\documentclass[11pt,a4paper]{article}

\usepackage{jheppub}
\usepackage{amsmath}
\usepackage{amssymb}
\usepackage{amsthm}
\usepackage{graphicx}
\usepackage{slashed}
\usepackage{setspace}
\usepackage{multicol}
\usepackage{hyperref}

\newcommand{\drawsquare}[2]{\hbox{%
\rule{#2pt}{#1pt}\hskip-#2pt
\rule{#1pt}{#2pt}\hskip-#1pt
\rule[#1pt]{#1pt}{#2pt}}\rule[#1pt]{#2pt}{#2pt}\hskip-#2pt
\rule{#2pt}{#1pt}}

\newcommand{\fund}{\raisebox{-.5pt}{\drawsquare{6.5}{0.4}}}
\newcommand{\Ysymm}{\raisebox{-.5pt}{\drawsquare{6.5}{0.4}}\hskip-0.4pt%
\raisebox{-.5pt}{\drawsquare{6.5}{0.4}}}
\newcommand{\Yasymm}{\raisebox{-3.5pt}{\drawsquare{6.5}{0.4}}\hskip-6.9pt%
\raisebox{3pt}{\drawsquare{6.5}{0.4}}}

\newcommand{\bZ}{\mathbb{Z}}

\newcommand{\cN}{\mathcal{N}}

\newcommand{\cO}{\mathcal{O}}

\newcommand{\ov}{\overline}

\newcommand{\eps}{\epsilon}

\newcommand{\OV}[1]{\cO_V^#1}
\newcommand{\OM}[1]{\cO_M^#1}
\newcommand{\OH}[1]{\cO_H^#1}

\title{Stringy Hidden Valleys}

\author[1,2]{Mirjam Cveti\v c,}
\author[1,3]{\hspace{.2cm}James Halverson,}
\author[1]{\hspace{.2cm}and Hernan Piragua} 
\affiliation{$^1$ Department of Physics and Astronomy, \\University of Pennsylvania,
  Philadelphia, PA 19104-6396, USA \vspace{.25cm} }
\affiliation{$^2$ Center for Applied Mathematics and Theoretical Physics,\\
University of Maribor, Maribor, Slovenia \vspace{.25cm} }
 \affiliation{$^3$ Kavli Institute for Theoretical Physics, \\ University of California,
  Santa Barbara, CA 93106-4030, USA \vspace{.25cm} }

\abstract{ We study gauge theories where quasi-hidden sectors are
  added to the MSSM for the sake of string consistency conditions
  which would otherwise not be satisfied. We focus on quiver gauge
  theories motivated by weakly coupled type II orientifold
  compactifications. Model independent features in this class include
  an anomalous $U(1)_V$ symmetry which protects messenger masses and
  has strong consequences for superpotential couplings, a rich
  phenomenology of heavy and light $Z'$ bosons, and axionic couplings
  required for anomaly cancellation via the Green-Schwarz
  mechanism. We discuss possibilities for dark matter and
  supersymmetry breaking in light of these generic features. Dark
  matter is necessarily non-baryonic, though many dark matter
  candidates have weak interactions. Most models have a $U(1)_VYY$
  anomaly whose cancellation requires couplings which allow for dark
  matter annihilation into photons through intermediate axions or
  anomalous $Z'$ bosons, as in two recently proposed scenarios. There is often
  an additional non-anomalous $U(1)$ symmetry which can give rise to a
  Fayet-like model of metastable supersymmetry breaking. Breaking of
  supersymmetry via SQCD can also be realized and flavor masses are often
  protected. Natural possibilities
  for mediation include gauge mediation, $Z'$ mediation, and
  D-instanton mediation, though it is not possible to realize minimal
  gauge mediation with messengers added for string consistency.  }

\begin{document}
\begin{flushright}
{\small \tt
  \tt UPR-1241-T \\
  \tt NSF-KITP-12-187
}
\end{flushright}
\maketitle

\section{Introduction}
With new experimental data arriving from the Large Hadron Collider and
other experiments, recent model building efforts include the study of
quasi-hidden gauge sectors which could be discovered in the near
future. Such sectors are well motivated from top-down constructions,
and it is possible that they interact weakly, but non-trivially, with
the standard model. These include hidden valleys
\cite{Strassler:2006im}, Higgs portals (for example \cite{Batell:2011pz}),
dark photons \cite{Jaeckel:2010ni}, and a variety of dark matter
models \cite{Cheung:2010gj}.  Hidden valleys, in particular, are strongly motivated by works on
string compactifications, such as \cite{Cleaver:1998sm,Cleaver:1998gc}. However, the top-down input into these
broad classes of effective field theories is that hidden extensions of
the standard model frequently exist, and therefore they should be
considered. It would be better, when possible, to have a guiding
principle from top-down considerations which lead to precise gauge theoretic
structures.

Gauge theories arising in string compactifications are constrained by
consistency conditions. For example, the ten dimensional type I
superstring is anomaly free only for gauge group
$\text{Spin}(32)/\bZ_2$. Global consistency conditions can also require the
presence of hidden sectors. In the heterotic $E_8 \times E_8$ string,
standard model sectors are typically realized in one $E_8$ factor,
while the other $E_8$ factor can give rise to rich hidden sector
physics which depend on consistency conditions on a holomorphic
vector bundles. In type II compactifications,
similar constraints are placed on the ranks of gauge groups by
Ramond-Ramond tadpole cancellation. These constraints can descend to
constraints on the chiral spectrum which are necessary for string
consistency.
The constraints on the chiral matter spectrum are not always
equivalent to the cancellation of four-dimensional gauge anomalies.
On one hand, string theory provides mechanisms for the cancellation of
certain anomalies, and therefore does not place constraints on the
chiral matter spectrum which ensure their absence. On the other
hand, there exist constraints on the chiral matter
spectrum of a gauge theory which are necessary for string consistency,
but for which there is no known field theoretic analog. We refer to
such constraints as ``stringy'' constraints.  

It is possible to enumerate all realizations of the MSSM in certain
classes of gauge theories motivated by string compactification.  Many
of these theories do not satisfy conditions necessary for string
consistency, and therefore some augmentation is required.  We focus
exclusively on the possibility of adding hidden sectors, where
messenger fields are chosen in a way that remedies visible sector
inconsistencies. Doing so can lead to specific gauge theoretic
structures which are generic across a broad class of models and have
important consequences for low energy physics. Extensions of the
MSSM with hidden sectors added for consistency form a proper
subset of the broader class of hidden valley models. For brevity, will
refer to a model in this class as a ``stringy hidden valley.''

\begin{figure}[t]
\centering
\includegraphics[scale=0.5]{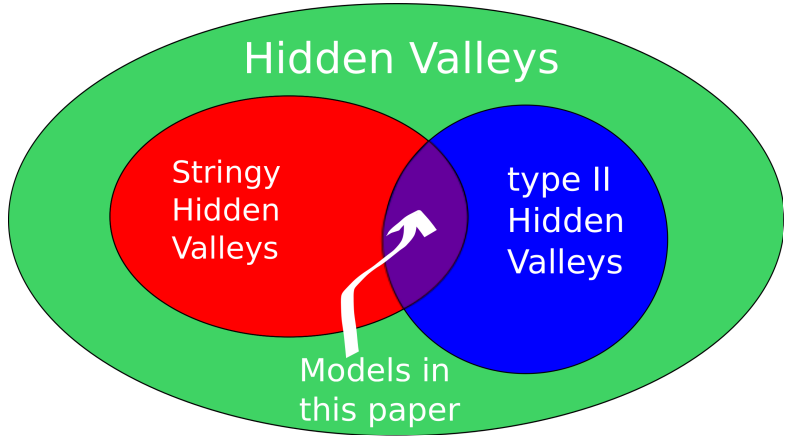} 
\caption{A depiction of different subsets of hidden valley models. The red region denotes
those models where hidden extensions of the standard model are added for the sake of
string consistency. The blue region denotes any hidden extension of a type II MSSM quiver.
In this paper we study hidden extensions of type II MSSM quivers which are added for
string consistency, denoted by the purple region.}
\label{figure:venn}
\end{figure}

In this paper we study a specific class of stringy hidden valleys.  In
a class of quiver gauge theories motivated by weakly coupled type II
orientifold compactifications and their duals, nearly all MSSM quivers
do not satisfy the constraints necessary for Ramond-Ramond tadpole
cancellation. The violation of these constraints is suggestive of
matter fields that should be added to the theory, and we study the
possibility that these fields are messengers to a hidden sector. We
will find that this class of theories generically has messenger masses
protected by symmetry, a rich structure of superpotential couplings,
axionic couplings required for anomaly cancellation, and $Z'$
bosons. These generic features give rise to interesting possibilities
for supersymmetry breaking and dark matter, including annihilation
processes with photons in the final state. We emphasize that these
features are non-generic within the broad class of hidden valley
models, but are generic in the class we study. We also emphasize that
our results do not depend strongly on the number of nodes in the
visible sector or possible flux contributions to tadpole cancellation
conditions, as we will
discuss. For brevity throughout, we will refer to these particular
models as stringy hidden valleys, keeping in mind that they are a
subset of stringy hidden valleys motivated by type II orientifold
compactifications. See figure \ref{figure:venn}.

Let us briefly describe the broader class of theories.
Quiver gauge theories arising in weakly coupled type II
orientifold compactifications\footnote{Note that the quivers
we consider  differ slightly from those arising from D3-branes at
  singularities. There, the singularity structure determines the
  quiver. Here our approach is from the bottom-up: we build MSSM
  quivers and hidden sector extensions using the ingredients of type
  II string theory, independent of any global embedding or realization
  at a particular singularity. For D3-brane quivers, see for example
  \cite{Aldazabal:2000sa,Verlinde:2005jr,Krippendorf:2010hj} and
  references therein.} and their duals have generic
properties which affect low energy physics, including the
presence of anomalous $U(1)$ symmetries which can forbid
superpotential couplings in perturbation theory. These terms can be
generated with suppression via D-instanton effects
\cite{Blumenhagen:2006xt,Ibanez:2006da,Florea:2006si} or via couplings
to singlets. Cancellation of $U(1)$ anomalies\footnote{This includes
  cubic abelian, mixed abelian non-abelian, and mixed
  abelian-gravitational anomalies.} via the Green-Schwarz mechanism
requires the presence of Chern-Simons couplings of the form $B \wedge
F$ and $\phi F \wedge F$, where the former gives a large St\"
uckelberg mass to the associated $Z'$ gauge boson. Consistent quivers
must also satisfy conditions necessary for Ramond-Ramond tadpole
cancellation. If they are to realize the MSSM, quivers must satisfy
constraints necessary for the hypercharge to remain massless.

There has been much phenomenologically motivated work studying type II
quiver gauge theories without hidden sectors. These studies have been both systematic and
example-driven. For example, it has been shown
\cite{Ibanez:2008my,Anastasopoulos:2009mr,Cvetic:2009yh,Cvetic:2009ez,Cvetic:2009ng,Anastasopoulos:2009nk}
that D-instantons in MSSM quivers can give rise to realistic Yukawa
couplings and neutrino masses while ensuring the absence of R-parity
violating operators and dimension five proton decay
operators. Alternative mechanisms for realistic neutrino masses have
been proposed, including the generation of a realistic Dirac neutrino
mass term \cite{Cvetic:2008hi} and a Weinberg operator $LH_uLH_u$
\cite{Cvetic:2010mm} by D-instantons.  Other issues which have been
studied include singlet-extended standard models \cite{Cvetic:2010dz},
dynamical supersymmetry breaking \cite{Fucito:2010dk}, grand
unification \cite{Kiritsis:2009sf,Anastasopoulos:2010hu}, and $Z'$
physics
\cite{Berenstein:2006pk,Cvetic:2011iq,Anchordoqui:2011eg}. Systematics
of hypercharge embeddings, bottom-up configurations, and global
rational conformal field theory realizations have been carried out in
\cite{Anastasopoulos:2006da}.

This paper is very similar in spirit to \cite{Cvetic:2011iq}, which also utilized string consistency conditions as the guiding
principle for physics beyond the standard model. Since it is so closely related to
this work, let us briefly review the conclusions. In \cite{Cvetic:2011iq}, exotic matter was
added to inconsistent type II MSSM quivers for the sake of
consistency and in the most general way allowed by the
construction, without adding additional quiver nodes. It was shown at the level of standard model
representations that some exotics are much more likely than others,
with a clear preference for MSSM singlets, $SU(2)_L$ triplets without
hypercharge, and quasichiral pairs\footnote{Quasichiral pairs are
  vector with respect to the standard model but chiral with respect to
  some other symmetry in the theory. In this case the symmetry is an
  anomalous $U(1)$.}. All of these possibilities could be relevant for
LHC physics, as mass terms for the quasichiral pairs are forbidden by
an anomalous $U(1)$ symmetry but could be generated at the TeV scale
by string instantons or couplings to singlets. We also refer
the reader to \cite{Cvetic:2011iq} for a more in depth discussion of the
constraints we will import on quivers.
In this paper, we will study the same class of MSSM quivers but will
add hidden sectors for the sake of consistency.

This paper is organized as follows. In section \ref{sec:basic setup}
we review the structure of the quiver gauge theories we study,
including string consistency conditions, and also the classification
of three-node MSSM quivers. We introduce the possibility of
adding hidden sectors only for the sake of string consistency and set
notation used throughout. In section \ref{sec:general physics} we show
that hidden sectors added for consistency have generic features which affect low
energy physics, including an important symmetry $U(1)_V$, axionic
couplings, and $Z'$ bosons. These have important implications for
various dark matter and supersymmetry breaking scenarios, which we
discuss in detail. In section \ref{sec:examples} we present explicit
quivers realizing the general ideas of section \ref{sec:general physics}.
In section \ref{sec:conclusions} we conclude, briefly stating the main
results and discussing possibilities for future work.

\section{Basic Setup and Guiding Principles}
In this section we will introduce a class of quiver gauge theories,
emphasizing how they can arise in weakly coupled type II string
compactifications.  We discuss constraints on their chiral matter
spectrum which are necessary for string consistency and introduce a
guiding principle for adding hidden sectors.  We will
review the classification of three-node MSSM quivers. We will also discuss the
applicability of our results to extensions of higher-node MSSM quivers.  

\label{sec:basic setup}

\subsection{Symmetries, Spectrum, and String Consistency}
\label{sec:symms spec and consistency}
Let us introduce the class of theories we consider\footnote{For a
  recent in-depth discussion of this class of theories, see
  \cite{Cvetic:2011vz}.}, beginning with a discussion of symmetries.
A compactification of weakly coupled type II string theory realizes
gauge degrees of freedom as open strings ending on D-branes. For
specificity, consider type IIa.  A D6-brane wrapped on a generic
three-cycle $\pi_i$ exhibits $U(N_i)$ gauge symmetry, though $SO(N_i)$
or $Sp(N_i)$ gauge symmetry can exist if $\pi_i$ happens to be an
orientifold invariant cycle. For generality and simplicity, we
consider all gauge factors to be $U(N_i)$. In the presence of
O6-planes, every D6-brane has an associated orientifold-image brane,
which will be important for the spectrum. We represent a brane
together with its image brane as a node in a quiver, labeled by
$U(N_a)$. The trace $U(1)$ of a $U(N_i)$ factor is often anomalous,
and the associated abelian and mixed anomalies are automatically
canceled in globally consistent string compactifications by
four-dimensional terms coming from the dimensional reduction of the
Chern-Simons D-brane action.  In a bottom-up
gauge theory, the necessary Chern-Simons terms can be added by
hand (see \cite{Anastasopoulos:2006cz}, e.g.). In particular, anomaly cancellation requires
terms of the form $\int
d^4x\,\, B \wedge F$, which gives a string scale St\" uckelberg mass
to the $U(1)$ with associated field strength $F$. Though these degrees of freedom can
be integrated out at scales well below the associated $Z'$ mass, the massive $U(1)$'s impose global selection rules
on the low energy effective action, forbidding many terms in the
superpotential. We will realize the standard model gauge group via
unitary factors $U(3)_a\times U(2)_b \times \prod_{\alpha=1}^N U(1)_\alpha$.  We
require the one linear combination $U(1)_Y = q_a U(1)_a + q_b U(1)_b +
\sum_{\alpha=1}^N q_\alpha U(1)_\alpha$ is left massless\footnote{Throughout, we will
  use the phrases ``massless U(1)'' or ``light $Z'$'' to describe a
  $U(1)$ symmetry which is not required to obtain a St\" uckelberg
  mass. The terms ``massive U(1)'' or ``heavy $Z'$'' will be used to
  denote a $U(1)$ symmetry which does not satisfy the constraints
  (\ref{eqn:chiral masslessness constraint}) and therefore obtains a
  string scale St\" uckelberg mass.} and can be identified as
hypercharge.  Such a combination is referred to as a hypercharge
embedding. In addition, it is possible that another linear combination
is left massless and is associated to a light $Z'$ which must obtain a
mass via the standard Higgs mechanism.

\begin{figure}[t]
\centering
\includegraphics[scale=0.7]{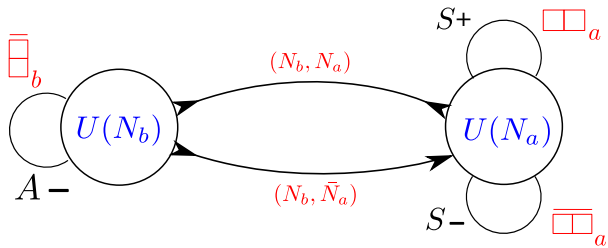} 
\caption{Example of a quiver. Each black line represents a field. The red symbols 
next to them are the associated representations, which we will henceforth omit since this data is equivalently communicated by the decorated edges. See the text for more details about this convention. }
\label{figure:notation-example}
\end{figure}

Let us discuss the spectrum.  Chiral matter is localized at the
intersection of two D6-branes and the chiral index is given
by the topological intersection number of two three-cycles. Given a D6-brane on
$\pi_a$, another D6-brane on $\pi_b$ and its image on $\pi_b'$,
strings localized at intersections of $\pi_a$ with $\pi_b$ carry
representations $(N_a,\ov N_b)$ or $(\ov N_a,N_b)$ under $U(N_a)\times
U(N_b)$ and strings localized at the intersections of $\pi_a$ and
$\pi_b'$ carry representations $(N_a,N_b)$ or $(\ov N_a,\ov N_b)$. The
fundamental and antifundamental representations carry charge $\pm 1$
under the trace $U(1)$'s, respectively.  Therefore the branes can
realize all four combinations of bifundamental representations, which
we present as a bidirectional edge between the $N_a$ node and the
$N_b$ node. String zero modes localized at the intersection of $\pi_a$
with $\pi_a'$ come in symmetric or antisymmetric tensor
representations under $U(N_a)$ and carry charge $\pm 2$ under the
trace $U(1)_a$ associated to $U(N_a)$. We label these representations
as an arrow from the $N_a$ node to itself with a $+$ or $-$ to denote
the sign of the $U(1)$ charge and $S$ or $A$ to denote symmetric or
antisymmetric. See figure \ref{figure:notation-example} for an example.
For a given set of nodes and hypercharge embedding, it is straightforward to enumerate 
all possible realizations of the MSSM spectrum.

We require that the chiral matter spectrum of the quiver satisfies two
sets of conditions. The first set are those necessary for tadpole
cancellation\footnote{Tadpole cancellation is the requirement that the
  net Ramond-Ramond (D-brane) charge is canceled on the internal
  space. It is necessary for the consistency of a globally defined
  string compactification. For example, in type IIa tadpole
  cancellation places a homological constraint on three-cycles wrapped
  by D6-branes which can be shown to descend to the weaker conditions
  (\ref{eqn:chiral tadpole constraint}) on chiral matter. These
  conditions are necessary but not sufficient for D6-brane tadpole
  cancellation. }, which are
\begin{align}
\label{eqn:chiral tadpole constraint}
N_a \ge 2&: \quad \# a - \# \ov a + (N_a+4)\,\, (\# \, \Ysymm_a - \#\, \ov \Ysymm_a) + (N_a-4) \,\, (\# \, \Yasymm_a - \# \, \ov \Yasymm_a) = 0 \notag \\ \notag \\
N_a = 1&: \quad \# a - \# \ov a + (N_a+4)\,\, (\# \, \Ysymm_a - \# \, \ov \Ysymm_a) = 0 \,\,\, \text{mod} \,\,\, 3,
\end{align}
for each $U(N_a)$ node, where we have denoted the fundamental and
antifundamental by $a$ and $\ov a$.  A $U(1)$ defined by an arbitrary
linear combination $\sum_x q_x \, U(1)_x$ will obtain a string scale
St\" uckelberg mass unless the masslessness conditions
\begin{align}
\label{eqn:chiral masslessness constraint}
& N_a \ge 2: \nonumber \\  & -q_aN_a\,\,(\#\Ysymm_a - \#\ov\Ysymm_a + \#\Yasymm_a - \#\ov\Yasymm_a) + \sum_{x\ne a} q_x N_x \,\, (\#(a,\ov x) + \#(\ov a, \ov x) - \#(\ov a, x) - \#(a,x)) = 0 \notag \\ 
& N_a = 1: \nonumber \\ & -q_a \,\,\frac{\#a - \#\ov a + 8(\#\Ysymm_a-\#\ov\Ysymm_a)}{3} + \sum_{x\ne a} q_x N_x \,\, (\#(a,\ov x) + \#(\ov a, \ov x) - \#(\ov a, x) - \#(a,x)) = 0.
\end{align}
are satisfied. The second set of conditions we impose is that the hypercharge embedding satisfies
these masslessness conditions. In a given quiver, there may also be other linear combinations which satisfy
these equations, giving light $Z'$ bosons.

Let us define some terminology that we will use throughout to discuss
these conditions.  For a $U(N_a)$ gauge node, we refer to the lefthand
side of the conditions necessary for tadpole cancellation and a
massless hypercharge boson as the ``T-charge'' $T_a$ and the
``M-charge'' $M_a$. In the three-node MSSM quivers we consider with
$U(3)_a\times U(2)_b \times U(1)_c$, we will ensure that the T-charges
$T_a,T_b,$ and $T_c$ satisfy \eqref{eqn:chiral tadpole constraint} and
the M-charges $M_a,M_b,$ and $M_c$ satisfy the equations \eqref{eqn:chiral
  masslessness constraint}. In addition, we may refer to the contributions
of certain sets of fields to some T-charge or M-charge, where the context
will make the content clear. For example, $T_b^{\text{mess}}$ could be
the contribution of messenger fields to $T_b$.

The equations necessary for tadpole cancellation for $N_a \ge 3$ are
the conditions necessary for the cancellation of cubic $SU(N_a)$
anomalies. It is crucial that consistent chiral spectra do not give
rise to these anomalies, since the Green-Schwarz mechanism cannot
cancel them. The corresponding field theory constraints do not exist
for $N_a=2$ or $N_a=1$, however, and we refer to these as ``stringy''
constraints.  We refer the reader to \cite{Cvetic:2011iq} for a recent
in-depth discussion of these constraints and field theoretic
constraints. These constraints are often violated for MSSM quivers. Our guiding principle will be to add hidden sectors so that they are satisfied.

\subsection{Classifying Three-node MSSM Quivers}
\label{sec:three node MSSM}
Our results regarding stringy hidden valleys will apply to
essentially any MSSM quiver with non-zero T-charge which is canceled by the
non-zero T-charge of messengers to a hidden sector. However,
three-node MSSM quivers and their extensions provide an
excellent example. Let us review their classification.
 
Consider a quiver with $U(3)_a\times U(2)_b \times U(1)_c$ gauge
symmetry, which is the minimal number of nodes that can realize the
MSSM gauge group and chiral spectrum at low energies. $SU(3)\times
SU(2)_L$ of the standard model arise from the $U(3)$ and $U(2)$
factors and hypercharge must arise as a linear combination
\begin{equation}
U(1)_Y = q_a \, U(1)_a + q_b \, U(1)_b + q_c \, U(1)_c
\end{equation} 
of the trace $U(1)$'s. There are only two possible
sets $(q_a,q_b,q_c)$ which can realize the entire MSSM spectrum utilizing
bifundamental, symmetric, and antisymmetric tensor representations. The
first is the well-known Madrid embedding \cite{Ibanez:2001nd}, given by
\begin{align}
\label{eqn:Madrid embedding}
U(1)_Y = \frac{1}{6} \, U(1)_a + \frac{1}{2} \, U(1)_c
\end{align}
and possible MSSM field representations given by\footnote{To avoid unnecessary notation throughout,
we make the definitions $u^c \equiv u_L^c$, $d^c \equiv d_L^c$, $e^c \equiv e_L^c$, and $\nu^c \equiv \nu_L^c$.}
\begin{align}
\label{eqn:Madrid MSSM fields}
 Q: \,\,\,\, (a,b) \,\,\, (a,\ov b), \qquad  &u^c : \,\,\,\, (\ov a, \ov c), \qquad  d^c: \,\,\,\, \Yasymm_a \,\,\, (\ov a,  c), \nonumber \\
 L: \,\,\,\, (b,\ov c) \,\,\, (\ov b, \ov c), \qquad & e^c : \,\,\,\, \Ysymm_c,  \nonumber  \\
 H_u: \,\,\,\, (b, c) \,\,\, (\ov b, c), \qquad  & H_d : \,\,\,\, (b, \ov c) \,\,\, (\ov b, \ov c),
\end{align}
where the unbarred and barred letters represent the fundamental and antifundamental representations of the associated nodes.
MSSM singlets can be realized as $\Yasymm_b$ or $\ov \Yasymm_b$, and for this embedding we define the chiral excess of
singlets to be $N_S \equiv \#\Yasymm_b - \#\ov \Yasymm_b$.
Lacking a better name, the other hypercharge embedding is the non-Madrid embedding, given
by
\begin{align}
\label{eqn:non-Madrid embedding}
U(1)_Y = -\frac{1}{3} \, U(1)_a - \frac{1}{2} \, U(1)_b
\end{align}
and the possible MSSM field representations are given by
\begin{align}
\label{eqn:non-Madrid MSSM fields}
Q: \,\,\,\, (a,\ov b), \qquad  & u^c: \,\,\,\, \Yasymm_a, \qquad  d^c: \,\,\,\, (\ov a, c) \,\,\, (\ov a,  \ov c), \nonumber \\
L: \,\,\,\, (b, c) \,\,\, (b, \ov c), \qquad & e^c: \,\,\,\, \ov \Yasymm_b, \nonumber \\
H_u: \,\,\,\, (\ov b, c) \,\,\, (\ov b, \ov c), \qquad &  H_d: \,\,\,\, (b, \ov c) \,\,\, (b, \ov c).
\end{align}
MSSM singlets can be realized as $\Ysymm_c$ or $\ov \Ysymm_c$, and for this embedding we
define the chiral excess
of singlets to be $N_S \equiv \# \Ysymm_c - \# \ov \Ysymm_c$.
Depending on the coupling of these singlets to MSSM fields, they could
be right-handed neutrinos $\nu^c$ or singlets $S$ which give rise to a
dynamical $\mu$-term. See \cite{Cvetic:2010dz} for singlet-extended MSSM quivers in this class.

Given the possible MSSM field representations in the Madrid and
non-Madrid embeddings, (\ref{eqn:Madrid MSSM fields}) and (\ref{eqn:non-Madrid MSSM fields}), it is
possible to enumerate\footnote{Since $L$ and $H_d$ carry the same standard
model quantum numbers, we treat them as one field with multiplicity four in our counting.} all three-node realizations of the exact MSSM
spectrum. One can also compute the T-charges and M-charges from
equations (\ref{eqn:chiral tadpole constraint}) and (\ref{eqn:chiral
  masslessness constraint}). The possible T-charges for the Madrid
quivers are given by
\begin{equation}
\label{eqn:Madrid T-charge M-charge}
T_a = 0 \qquad T_b = \pm 2n \qquad T_c=0\, \text{mod}\, 3 \qquad\qquad \text{with} \qquad n \in \{0,\dots, 7 \}
\end{equation}
and all M-charges are automatically zero. That is, all of the conditions necessary for a massless hypercharge
are satisfied and the conditions necessary for tadpole cancellation are only violated on the $U(2)_b$ node. If a chiral excess $N_S$ of singlets are added to the theory, the only difference is $T_b = \pm 2n - 2N_S$. Performing the same analysis for the non-Madrid embedding, one obtains
\begin{equation}
\label{eqn:non-Madrid T-charge M-charge}
T_a = 0 \qquad T_b = 0 \qquad T_c=\{0,1,2\} \,\, \text{mod} \,\, 3,
\end{equation}
and all M-charges zero. Therefore, for the non-Madrid embedding the
only T-charge or M-charge violation is on the $U(1)_c$ node. If a
chiral excess $N_S$ of singlets are added to the theory, the possible
T-charges and M-charges remain the same but the multiplicity of
quivers changes due to the new fields. For the Madrid embedding with
$N_S=0$, there are 160 quivers in all, 144 of which violate the $T_b$
condition. For the non-Madrid embedding with $N_S=0$, there are 40
quivers in all, 24 of which violate the $T_c$ condition. With regards
to anomalies, a simple calculation shows that 144 of the $160$ Madrid
quivers have a $U(1)_bYY$ anomaly, and all non-zero anomaly
coefficients are non-integral. All $40$ quivers with the non-Madrid
quivers have a $U(1)_cYY$ anomaly, and all anomaly coefficients are
non-integral. We will utilize these facts later when discussing hidden
sectors.

It is remarkable that \emph{most} three-node MSSM quivers violate the
conditions necessary for tadpole cancellation.  This is also true of MSSM quivers
with a larger number of nodes. Let us briefly discuss
possible field theoretic explanations in terms of anomalies. Since the spectrum 
under $SU(3)\times SU(2)_L
\times U(1)_Y$ of each quiver is the exact MSSM, perhaps with
singlets added, there are no cubic
non-abelian anomalies and there is no global $SU(2)$ anomaly
\cite{Witten:1982fp}.  Any mixed anomalies involving abelian
symmetries, such as the $U(1)_bYY$ and $U(1)_cYY$ anomalies
just discussed, can be canceled by the introduction of appropriate
Chern-Simons terms\footnote{Consistent type II string
  compactifications provide these terms automatically via dimensional
  reduction of the Chern-Simons D-brane action. See
  \cite{Anastasopoulos:2006cz}, for example, for a similar discussion
  purely in field theory.}, and therefore these quivers are consistent
quantum field theories. However, without further modification any
quiver which does not satisfy the conditions necessary for tadpole
cancellation cannot be embedded into the types of string
compactifications we have discussed.  One possible solution is to add
matter to the theory so that the inconsistent quivers become
consistent. This was done for three-node quivers in
\cite{Cvetic:2011iq}, as discussed in the introduction, where it was
shown that string consistency conditions prefer some standard
model representations over others. We now turn to another
possible solution.

\subsection{Adding Hidden Sectors for Consistency}
\label{sec:hidden sectors sec 2}
In section \ref{sec:three node MSSM}, we showed that most three-node
quivers with the exact MSSM spectrum are not consistent with
conditions necessary for tadpole cancellation, despite being
consistent as quantum field theories after the addition of
Chern-Simons terms.  Higher node MSSM quivers also exhibit this
behavior.  We discussed the possibility of adding matter to the
visible sector nodes to cancel the T-charge contribution of MSSM
fields.

\begin{figure}[t]
\centering
\includegraphics[scale=0.6]{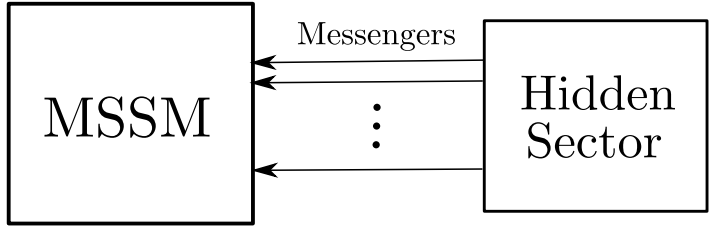} 
\caption{Heuristic depiction of the stringy hidden valleys studied in this paper. The setup is similar to standard hidden sector setups,
but the messengers are chiral under an anomalous $U(1)_V$ symmetry in the visible sector in order to cancel the non-zero $T_V$ charge of MSSM fields. }
\label{figure:mssm-m-h}
\end{figure}

Another possible solution is to add hidden gauge sectors where
bifundamental messenger fields with one end on a visible sector node
$V$ cancel any overshooting in T-charge or M-charge, ensuring that
hidden sector is also consistent. This will be our guiding principle
for physics beyond the standard model. We call\footnote{As emphasized
  in the introduction, stringy hidden valleys are potentially a much
  broader class than the models studied here. More specifically, we
  are studying stringy hidden valleys in type II quivers.} any hidden
sector of this type a ``stringy hidden valley.'' The setup is
heuristically depicted in figure \ref{figure:mssm-m-h} and looks
similar to common depictions of hidden sectors. We have a visible
sector with fields transforming under $SU(3)\times SU(2)_L\times
U(1)_Y$ and associated anomalous $U(1)$'s, a hidden sector with fields
transforming under some hidden sector gauge group $G_H$, and messenger
fields transforming under both the visible sector group and
$G_H$. However, the setups we study will differ from generic hidden
sectors in at least two important ways. First, since our hidden
sectors are added to cancel some visible sector $T_V$-charge, the
messenger fields to the $V$ node will always vector-like with respect to the standard
model but chiral\footnote{We will refer to such fields
  as ``quasichiral''.} under $U(1)_V$,  and therefore their masses are always
protected. Phenomenologically this very important, since pairs which
are vector with respect to \emph{all} symmetries have string scale
masses at a generic point in the moduli space of a string
compactification. Second, the structure of the hidden sectors will be
constrained by string consistency conditions.  

In the three-node
Madrid and non-Madrid embeddings string consistency requires that
messengers are added to the $U(2)_b$ and $U(1)_c$ nodes, respectively,
and therefore the symmetry $U(1)_V$ which charges the messengers are
the trace $U(1)_b$ and $U(1)_c$. In discussing $U(1)_V$, though, we
will be able to address aspects of low energy physics in these models
which do not depend on the visible sector hypercharge embedding.

There are two basic types of nodes in a hidden sector: those connected directly
to the visible sector via messengers and those which are not. We utilize
lower case latin indices for the first type, calling them $H_i$ nodes,
and capital latin indices for the second type, calling them $H_I$ nodes.
The most general hidden sector could be composed of $n$ connected graphs,
each disjoint from one another but connected to the visible sector
via messengers. When differentiating between disconnected clusters in
the hidden sector, we will utilize a superscripted $(m)$ to describe
quantities in the $m^{th}$ cluster. For example, $H_I^{(m)}$ nodes would
be nodes in the $m^{th}$ cluster which are not directly connected to the visible
sector.

\subsubsection{``Hidden Hypercharge'' Quivers}

In adding a hidden sector, it is possible that the hypercharge
embedding is modified due to contributions from trace $U(1)$'s of
hidden sector nodes. Before adding hidden sectors, the MSSM quivers
had a hypercharge embedding given by a linear combination of trace
$U(1)$'s of visible sector nodes. We previously called this $U(1)_Y$,
but henceforth will call it $U(1)_{Y,V}$. We write the full
hypercharge embedding as a linear combination
\begin{equation}
U(1)_Y = U(1)_{Y,V} + U(1)_{Y,H},
\end{equation}
where $U(1)_{Y,H}$ is the contribution from hidden sector nodes.
While it may seem strange to discuss hidden sector contributions to
hypercharge, it is possible to nevertheless ensure that all hidden
sector fields are MSSM singlets, and thus we should consider this possibility. We will
argue in a moment that it is actually generic to have $U(1)_{Y,H}$ non-trivial.
 We call any quiver with a non-trivial $U(1)_{Y,H}$ a
``hidden hypercharge'' quiver. We call its hidden sector a ``hypercharged
stringy hidden valley''.

Any hidden hypercharge quiver is very constrained. Consider the possibility of a
single-cluster hidden sector with non-trivial $U(1)_{Y.H}$. In general, we could have
\begin{equation}
U(1)_{Y,H} = \sum_\alpha q_\alpha \, U(1)_\alpha
\end{equation}
where the sum is over all hidden sector nodes indexed by $\alpha$.
For the cluster to be connected and hidden, however, every hidden sector node
must be connected to another hidden sector node by a field which is an
MSSM singlet.  This requires $|q_\alpha| = |q_\beta|$ for all $\alpha$
and $\beta$ where the type of bifundamental is dictated by the signs
of $q_\alpha$. We take $q_\alpha = q_\beta\equiv q$ for all
$\alpha,\beta$ without loss of generality, which requires that we use
only bifundamentals $(\alpha,\ov \beta)$ and $(\ov \alpha, \beta)$ and
not $(\alpha, \beta)$ or $(\ov \alpha, \ov \beta)$. Thus, the
non-trivial contribution to hypercharge ``propagates'' through the
entire cluster by the requirement that bifundamental fields are MSSM
singlets. For an $n$-cluster hidden sector, then, we have\footnote{We
  have defined $U(1)_F^{(m)}$ for later convenience, since it is a
  particularly natural linear combination to consider. We will see it can play a role
in both supersymmetry breaking and stabilization of messenger dark matter.}
\begin{equation}
U(1)_{Y,H} = \sum_{m=1}^n \,\,U(1)_{Y,H}^{(m)} \qquad \text{where} \qquad U(1)_{Y,H}^{(m)} \equiv q^{(m)} \sum_\alpha \,\,\, U(1)^{(m)}_\alpha \equiv q^{(m)} U(1)_F^{(m)}
\label{eqn:U1yh and U1myh}
\end{equation}
and the first sum is over the $n$ clusters while the second is over all nodes
in the $m^{th}$ cluster. The key observation is that all nodes in a given cluster
contribute to the hypercharge embedding in the same way. Thus, any cluster is labeled
by a rational number $q^{(m)}$ which determines its contribution to hypercharge. Physically,
$q^{(m)}$ has a simple interpretation: $\pm q^{(m)}$ is the hypercharge of the messengers to
the $m^{th}$ hidden cluster.

\begin{figure}[t]
\centering
\includegraphics[scale=0.7]{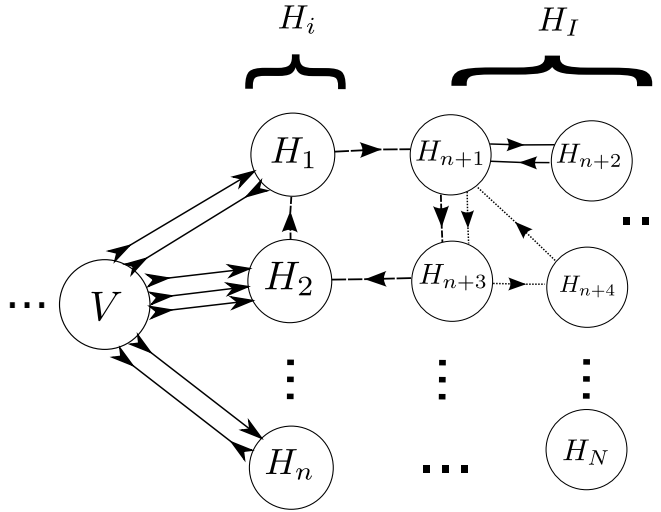} 
\caption{An example of a hidden sector. The first column of
  nodes, $H_i$, are connected to the visible sector node $V$ through the
  messengers. The nodes on the right $H_I$ do not have messengers attached. 
Closed loops, such as the ones given by the dotted and dashed lines,
give perturbative superpotential couplings. We have not labeled the
gauge groups to emphasize the general structure. However, $T_{H_2} \ne 0$
in this quiver requires that it is a $U(1)$ node. $H_1$ could be non-abelian.}
\label{figure:generic abelian madrid}
\end{figure}

At the level of graph theory, a hidden cluster with $q^{(m)}\ne 0$ is
a directed graph: any non-messenger edge is between two hidden nodes
and has its two arrows in the same direction, in which case we may
draw a single arrow indicating the direction.  Any loop in hidden
sector edges is directed, in contrast with the general case, and the
corresponding fields give a perturbative superpotential coupling
composed of MSSM singlets. For example, consider figure
\ref{figure:generic abelian madrid}.  The messengers are the fields
between nodes of type $H_i$ and node $V$, which has an anomalous $U(1)_V$
that charges the messengers. The hidden sector fields are
fields connecting hidden sector nodes. Letting $S_{ij}$ be the singlet
from nodes $i$ to $j$, one perturbative superpotential coupling in
this quiver is given by $\frac{1}{M_s} S_{1,n+1} S_{n+1,n+3} S_{n+3,2}
S_{2,1}$, represented by the closed loop between those nodes.

Since the notion of a hidden hypercharge quiver may still seem strange, let us make some comments regarding generality. We have argued that the requirement
that hidden sector fields are MSSM singlets fixes the hypercharge contribution of the $m^\text{th}$ hidden cluster
up to a single number $q^{(m)}$. For an $n$ cluster hidden sector, the contribution to the
hypercharge is determined by the tuple $(q^{(1)},q^{(2)}, \dots , q^{(n)})$. Only one possibility, given by the tuple $(0,\dots,0)$,
has a trivial contribution to hypercharge. Any other tuple gives a hidden hypercharge quiver. Though hidden hypercharge
quivers are not required by string consistency,
it would introduce a loss of generality to not consider them. Furthermore, we will argue in section \ref{sec:fchamp} that hidden
sector contributions to the hypercharge embedding are often required to avoid messenger fields with fractional electric charge.

\subsection{Fractionally Charged Massive Particles}
\label{sec:fchamp}
In this brief subsection we will address an important aspect of phenomenology
which must be taken into account when building quivers with hidden sectors.

Globally consistent string compactifications and quivers in the
class we study often exhibit particles with fractional electric
charge\footnote{Standard model quarks have fractional electric
charge, but this does not pose an issue since they are bound into
mesons and baryons, which have integral electric charge.}. If they exist, the lightest fractionally charged
massive particle is stable and its relic density is subject to strong
constraints from primordial nucleosynthesis and the cosmic microwave
background. Recent work \cite{Langacker:2011db} shows that their
existence is essentially ruled out. Therefore, we do not
consider quivers which give rise to particles with fractional electric charge.  This is an important
phenomenological consideration which greatly constrains the quivers we study.
For example, in extensions of the Madrid embedding any cluster with
a $U(1)$ $H_i$ type node must have $q^{(m)} \in \bZ + \frac{1}{2}$,
since the messengers are necessarily doublets of $SU(2)_L$ and otherwise they
would have fractional electric charge. In non-Madrid extensions, the
the messengers are necessarily singlets of $SU(3)_{QCD}\times SU(2)_L$ and therefore
any cluster with a $U(1)$ $H_i$ type node must have $q^{(m)} \in \bZ$. 
In a cluster with only non-abelian $H_i$ type nodes, confinement can relax these constraints,
though others exist which depend on the rank of the non-abelian gauge nodes.

\subsection{Comments on Fluxes and Consistency Conditions}

In the type IIb string, fluxes are crucial for moduli
stabilization and can contribute to tadpole cancellation
conditions. This is well known for the D3-tadpole, for example.  Since
the string consistency conditions (\ref{eqn:chiral tadpole
  constraint}) are necessary for tadpole cancellation, let us consider
how the conditions may change in the presence of fluxes.

We will work in type IIa, since the consistency conditions are easily
derived there. In the absence of fluxes, D6-brane tadpole
cancellation gives a homological constraint on the three-cycles wrapped by
D6-branes and O6-planes,
\begin{equation}
\label{eqn:fluxless tadpole}
 \sum_b N_b(\pi_b + \pi_{b'}) = 4 \pi_{O6}.
\end{equation}
The conditions (\ref{eqn:chiral tadpole constraint}) arise from intersecting
this equation with each cycle $\pi_\alpha$ on which a D6-brane stack is wrapped and utilizing
the relations between topological intersection numbers and chiral indices. See \cite{Cvetic:2011vz} for more details. The equations (\ref{eqn:chiral tadpole constraint}) can only be altered
if \eqref{eqn:fluxless tadpole} is altered. Schematically\footnote{See \cite{Kachru:2004jr,Camara:2005dc} for a detailed discussion of possible flux contributions to the D6-brane
tadpole.} this must be of the form
\begin{equation}
\label{eqn:fluxed tadpole}
 \pi_{\text{flux}} + \sum_b N_b(\pi_b + \pi_{b'}) = 4 \pi_{O6}.
\end{equation}
Intersecting with $\pi_\alpha$ gives a set of constraints similar (\ref{eqn:chiral tadpole constraint}), except for an additional possible contribution $T_\alpha^\text{flux} \equiv \pi_\alpha \cdot \pi_\text{flux}$. If $T_\alpha^\text{flux}=0$, the constraint on $T_\alpha$ is unchanged.
$T_\alpha^\text{flux}\ne 0$ the equations can be altered, though the flux must be tuned if it
is to precisely cancel any net T-charge of an MSSM quiver. In the generic case the flux
contribution will not exactly cancel the net T-charge of the MSSM quiver, 
and additional matter must still be added for the sake of consistency.

Therefore, the addition of fluxes will not significantly alter the physical conclusions of this
paper, which rely entirely on the fact that there is some net T-charge which is canceled
by quasichiral messengers to a hidden sector.

\section{General Structure of Low Energy Physics}
\label{sec:general physics}
Equipped with a guiding principle for adding hidden sectors, it is
possible to make statements about low energy physics which are true of
the stringy hidden valleys we consider, but not of a generic hidden
valley.  The major conclusions in this section will not require the specification
 of the visible sector matter content, the hidden sector matter content, or the
hypercharge embedding.   This is because stringy
hidden valleys generically exhibit symmetries and chiral spectra which
are non-generic within the class of all hidden valleys.  Presentation
of concrete quivers will be saved for section \ref{sec:examples}.

We will begin with a discussion of the symmetry $U(1)_V$ which charges
the messengers and the associated implications for superpotential
couplings.  Certain classes of couplings are forbidden in perturbation
theory; others are always highly suppressed. We will then turn to a
discussion of anomalies,  phenomenologically relevant
axionic couplings necessary for their cancellation, and $Z'$ physics.
We will show that these basic building blocks lead to interesting models
of dark matter and supersymmetry breaking, realizing mechanisms already
present in the literature.

\subsection{Light Messengers and Constrained Superpotential Couplings}
\label{sec:couplings}

One type of symmetry plays a distinguished role in all stringy hidden valleys.
Messengers are added with a net excess of $T_V$
charge for some visible sector node $V$, in which case they are chiral
under $U(1)_V$. In extensions of three-node quivers with
the Madrid or non-Madrid hypercharge embedding the node $V$ is the
$U(2)_b$ node or the $U(1)_c$ node, respectively, in which case messengers
are chiral under $U(1)_b$ or $U(1)_c$.  In the case of multiple visible
sector nodes $V$ which charge messengers of different type, there will be more than
one symmetry $U(1)_V$. We emphasize that the conclusions
of this section are generic for the models we study.

Let us discuss possible superpotential couplings, beginning with
couplings present in perturbation theory. Label a generic
superpotential coupling of $i$ chiral supermultiplets in the visible,
messenger, and hidden sectors as $\OV{i}$, $\OM{i}$, and $\OH{i}$,
respectively. Depending on the fields present in the coupling, it is
possible that $\OV{i}$ and $\OH{i}$ are present in perturbation
theory.  However, messenger fields carry anomalous $U(1)$ charge
of the same sign and therefore no operator $\OM{i}$ is present in
perturbation theory. This is particularly important for the messenger
mass terms $\OM{2}$, since the anomalous $U(1)$ symmetry prevents the
associated fields from acquiring a string scale mass and decoupling
from low energy physics. On general grounds, therefore, the messenger
mass is always protected by symmetry and
could be generated at the TeV scale via instantons or couplings to
singlet VEVs.  This is certainly not required in a generic hidden
valley or hidden sector, and it will have important consequences.

Symmetries also dictate the structure of mixed couplings
$\OV{i}\OM{j}\OH{k}$.  Since messenger fields are the only fields
transforming under both visible sector and hidden sector gauge nodes,
there are no closed paths in the quiver corresponding to couplings of
the form $\OV{i}\OH{j}$, and thus these operators are not present in
perturbation theory. Couplings of the form $\OV{i}\OM{1}\OH{j}$ are
forbidden since they carry always anomalous $U(1)$ charge. Thus, the
lowest dimension perturbative superpotential coupling involving both
visible sector and hidden sector fields is a non-renormalizable term
$\sim \frac{1}{M_s} \, \OV{1}\OM{2}\OH{1}$.  It is also possible to
couple messenger fields only to visible sector fields, and the lowest
dimension couplings of this type are $\sim \OV{1} \OM{2}$. Couplings
of messengers to hidden sector fields carry $U(1)_V$ charge and are
forbidden. We have exhausted the possibilities for couplings present
in perturbation theory.

Let us discuss couplings not present in perturbation theory, due to
carrying anomalous $U(1)$ charge. These coupling can be generated
non-perturbatively via D-instantons, in which case they are
exponentially suppressed by a factor $e^{-T}$, a suppression factor
dependent upon the volume of the cycle wrapped by the instanton.  A
perturbatively forbidden coupling $\cO$ can also be generated from a
perturbative couplings to singlets $\cO_S \sim S^n \cO$ if the singlet
has a vacuum expectation value.  Compared to the possibility of
obtaining $\cO$ directly in perturbation theory, obtaining it via
couplings to singlets suppress $\cO$ by a factor of $(\langle S \rangle / M_s)^n$. In either
case, the coupling $\cO$ receives a large suppression, which could\footnote{See \cite{Cvetic:2008hi}, for example, where instantons
  were used to generate Dirac neutrino masses of the observed order
  without the seesaw mechanism.} easily be $10^{-14}$.  The results for the
minimum suppression of a coupling $\cO$ are presented in table
\ref{table:operators-suppressions}, where we have utilized MSSM gauge
invariance and the chirality of messengers under $U(1)_V$ to
determine the minimum suppression for each coupling. For simplicity, we also require
that singlets which generate couplings are not messengers, since then $\cO_V\cO_H$
couplings could arise from $\cO_V \cO_M \cO_H$ couplings, for example. Relaxing
this assumption would complicate the analysis without significantly changing
the structure of couplings.

In summary, the structure of superpotential couplings is strongly constrained
by symmetries generically present in stringy hidden valleys. The most important observation compared to a generic quiver is that
$\OM{i}$ and $\OM{i} \OH{j}$ couplings are forbidden in perturbation theory,
and thus are very suppressed if they are present at all.
Suppression of these and other couplings play an important role in ensuring
that the hidden sector doesn't couple strongly to visible sector fields,
despite being present in the same connected quiver.

\begin{table}
\centering
\begin{tabular}{c |c c c|}
Coupling Structure & Perturbative & Instanton-induced & Singlet Couplings \\ \hline
$\OV{i}$ & $M_s^{3-i}$ & $e^{-T}M_s^{3-i}$ & $M_s^{3-(n+i)}$ \\ \hline
$\OM{i}$ & Forbidden & $e^{-T}M_s^{3-i}$ & $M_s^{3-(n+i)}$ \\  \hline
$\OH{i}$ & $M_s^{3-i}$ & $e^{-T}M_s^{3-i}$ & $M_s^{3-(n+i)}$ \\ \hline
$\OV{i} \OM{j} \OH{k}$ & $M_s^{3-(i+j+k)}$, $j\ge2$ & $e^{-T}M_s^{3-(i+j+k)}$  & $M_s^{3-(i+j+k+n)}$ \\ \hline
$\OV{i}\OM{j}$ & $M_s^{3-(i+j)}$ , $j\ge2$  & $e^{-T}M_s^{3-(i+j)}$ & $M_s^{3-(i+j+n)}$\\ \hline
$\OM{i}\OH{j}$ & Forbidden& $e^{-T}M_s^{3-(i+j)}$  & $M_s^{3-(i+j+n)}$ \\  \hline
$\OV{i} \OH{j}$ & Forbidden& $e^{-T}M_s^{3-(i+j)}$ & Forbidden \\ \hline
\end{tabular}
\caption{Couplings and their respective suppressions. All indices are $\ge 1$.  In the last column $n$ is the power of the singlet that
  acquires the VEV. In extensions of an MSSM quiver with the Madrid embedding there is extra suppression,
  since messengers transform under $SU(2)_L$ and gauge invariance requires at least $\OM{2}$ for
  couplings of type $\OM{i}$ and $\OM{i}\OH{j}$ obtained via instantons or singlet couplings. If
  the visible sector realizes the exact MSSM spectrum, then some couplings will be further suppressed by
  the requirement of MSSM gauge invariance.
  For example $\OV{i}$  and $\OV{i}\OM{j}$ would have $i \ge 2$.}
  \label{table:operators-suppressions}
\end{table}

\subsection{Anomalies, Required Axionic Couplings, and $Z'$ Bosons}
\label{subsec:z' physics}
All quivers in the broad class motivated by type II orientifold compactifications
have a rich structure of anomalies, axionic couplings required for their cancellation,
and $Z'$ physics. In this section we will make statements about them which are generic
for stringy hidden valleys, but not for the broader class.
We will see that these features can have important consequences for dark matter and
supersymmetry breaking. 

\subsubsection{For All Stringy Hidden Valleys}
We have emphasized that all stringy hidden sectors have messengers
which are chiral under an anomalous $U(1)$ symmetry $U(1)_V$.
Therefore there is a $U(1)_V U(1)_iU(1)_i$ anomaly for the $U(1)$ of
any $H_i$ node with $U(N_i)$ gauge symmetry. In addition, if $N_i > 2$,
there are mixed $U(1)_VSU(N_i)^2$ anomalies.

These anomalies must be canceled via the Green-Schwarz mechanism through
the introduction of Chern-Simons terms. The mixed abelian
anomalies are canceled by the introduction of terms of the form
\begin{equation}
\int d^4x \,\, B_V \wedge F_V \qquad \qquad \text{and} \qquad \qquad \int d^4x \,\, \phi_V \, F_i \wedge F_i,
\end{equation}
where $F_V$ and $F_i$ are the field strengths of $U(1)_V$ and $U(1)_i$, $B_V$ is the two-form
which gives a St\" uckelberg mass to $U(1)_V$, and $\phi_V$ is the zero-form which is the
four-dimensional Hodge dual of $B_V$, i.e. $dB_V = *_{4d}\,\,d \phi_V$. If
$N_i > 1$, the mixed abelian non-abelian anomalies require terms of the form
\begin{equation}
\int d^4x \,\, \phi_V \,\, Tr(G_i \wedge G_i),
\end{equation}
where $G_i$ is the field strength of $SU(N_i)$. These conclusions hold
for any stringy hidden valley, and the axionic terms
can play an important phenomenological role.

Since the coupling $\int B_V \wedge F_V$ is always present, $U(1)_V$ is always
a massive $U(1)$. $U(1)_V$ gauge interactions are suppressed via the large $Z'_V$
mass.  This has strong implications for interactions between messengers and visible
sector particles charged under $U(1)_V$. See section \ref{sec:z' mass} for a discussion
of $Z'$ masses and low energy physics. 

Finally, a non-generic but common possibility is to have a chiral excess of messengers\footnote{By this
  we simply mean that the net $U(1)_i$ charge of the messengers is non-zero.} on
the $H_i$ node. In this case there is a $U(1)_iU(1)_VU(1)_V$ anomaly, whose cancellation requires
terms of the form $\int B_i \wedge F_i$ and $\int \phi_i F_V \wedge F_V$ and $U(1)_i$
is a massive $U(1)$. If there is not an excess of chiral messengers on $H_i$, it is
possible that $U(1)_i$ is massless, but this cannot be determined without further
specification of the hidden sector spectrum. 

\subsubsection{For All Hidden Hypercharge Quivers}
In this section we discuss further aspects of anomalies, axionic couplings, and $Z'$
physics which are true of any hidden hypercharge quiver. The additional physics
is governed by the fact that the hidden sector contributes non-trivially to the 
hypercharge embedding, which will have interesting implications for dark matter
annihilation processes with photons in the final state.

\vspace{.5cm}
\noindent \textbf{\emph{$U(1)_VYY$ Anomalies and Axionic Couplings}}
\vspace{.3cm}

Let us first consider $U(1)_VYY$ anomalies. Since messengers carry hypercharge
and are chiral under $U(1)_V$, they will always contribute to this anomaly. 
Unlike $U(1)_VU(1)_iU(1)_i$ anomalies, $U(1)_VYY$ anomalies can also receive
contributions from visible sector fields, in which case it may be possible that
visible sector and messenger contributions cancel and there is no anomaly. We will
now argue that this is almost never the case for extensions of three-node quivers.
The only possible loophole is in the case where confinement relaxes constraints on
$q^{(m)}$ necessary to ensure the absence of fractionally charged massive particles.

Let us begin with the Madrid embedding where $U(1)_V = U(1)_b$, taking
$k$ messengers. As discussed in section \ref{sec:three node MSSM} all
but sixteen of the three-node MSSM quivers have a $U(1)_bYY$ anomaly,
and any non-zero anomaly coefficient is non-integral.  The $k$
messengers are $SU(2)_L$ doublets and give a net contribution to the
$U(1)_bYY$ anomaly of $\sum_m-2k(q^{(m)})^2$.  We must have 
$q^{(m)}=\frac{N_m}{2}$ with $N_m \in \bZ$ for either the
absence of fractionally charged massive particles\footnote{In a cluster with
a $U(1)$ node. As mentioned in section \ref{sec:fchamp} this constraint can
be relaxed if all nodes in the cluster $m$ are non-abelian.} or the existence of
an electrically neutral messenger component, and in this case, since
cancellation of $T_b$ charge requires $k$ even, the messengers give an
integral contribution to the $U(1)_bYY$ anomaly coefficient. Since the
contribution of all possible visible sectors to the anomaly
coefficient is non-integral, $\emph{all}$ stringy hidden valley
extensions of three-node MSSM quivers with the Madrid embedding and $q^{(m)} = \frac{N_m}{2}$
exhibit a $U(1)_bYY$ anomaly.  For non-Madrid extensions, we will often
require $q^{(m)}\in \bZ$ to ensure the absence of fractionally charged
massive particles, in
which case messengers will give an integral contribution to the
$U(1)_cYY$ anomaly which will not cancel the non-integral contribution
from visible sector fields. Thus, all non-Madrid quivers of this type
have a $U(1)_cYY$ anomaly. Summarizing, \emph{any} extensions of a
Madrid (non-Madrid) quiver with $q^{(m)}\in \bZ+\frac{1}{2}$ ($q^{(m)}
\in \bZ$) has a $U(1)_bYY$ ($U(1)_cYY$) anomaly.  

We have just argued that $U(1)_VYY$ anomalies are extremely common in hidden
hypercharge quivers, being very precise with the case of extensions of visible
sectors with three-nodes. For any quiver with such an anomaly, 
anomaly cancellation via the Green-Schwarz mechanism requires the presence of a term
\begin{equation}
\int d^4x \,\, \phi_V \,\, F_Y \wedge F_Y
\end{equation}
where $F_Y$ is the field strength of hypercharge. In section \ref{subsec:dark matter} will see
that these terms play a crucial role in recent models for dark matter
annihilation processes with photons in the final state.

\vspace{.5cm}
\noindent \textbf{\emph{The $Z'$ Physics of Hidden Hypercharge Quivers}}
\vspace{.3cm}

$Z'$ bosons appear in many top-down constructions and can greatly impact low
energy physics, as reviewed in \cite{Langacker:2008yv}.
The models we have proposed have a rich structure of $Z'$ physics. 
In hidden hypercharge quivers there is
a ``natural'' $U(1)$ which is usually massless, and is always
massless for any extension of a three-node quiver.  Recall from section
\ref{sec:hidden sectors sec 2} that we write the hypercharge embedding
as
\begin{equation}
U(1)_Y  = U(1)_{Y,V} + U(1)_{Y,H},
\end{equation}
where the two terms are the contributions of the visible sector and
the hidden sector to the hypercharge linear combination,
respectively. For all quivers, we require that the linear combination
$U(1)_Y$ satisfies the linear equations (\ref{eqn:chiral masslessness
  constraint}). If $U(1)_{Y,V}$ independently satisfies these linear
equations, as it does for all extensions of three-node MSSM
quivers\footnote{$U(1)_{Y,V}$ is just the Madrid or non-Madrid linear
  combination. These linear combinations are massless for all
  three-node MSSM quivers, and it is easy to see that adding a hidden
  sector will not cause these conditions to be violated for
  $U(1)_{Y,V}$.}, then $U(1)_{Y,H}$ will also satisfy these
equations. Therefore any such quiver will give rise to a light $Z'$
boson. We will think of this $Z'$ as coming from $U(1)_{Y,H}$ and
henceforth call it $Z'_{Y,H}$, though we could equivalently consider
$U(1)_{Y,V}$.  $Z'_{Y,H}$ couples to any messengers which have
hypercharge, but never to hidden sector fields. $U(1)_{Y,H}$ is
closely related to $U(1)_F$. If there is a single
cluster, $U(1)_{Y,H}$ is just $U(1)_F$ rescaled by $q\equiv q^{(m)}$.

There are further interesting statements that one can make about $Z'$
physics in hidden hypercharge quivers.  To do so, it is
useful to consider two possibilities for the cluster $m$: the case
where cluster $m$ has a chiral excess of messengers on some node $H_i^{(m)}$, and the case where it
has no such chiral excess for \emph{any} $H_i^{(m)}$ node.

Let us
first consider the possibility where there is no chiral excess of messengers
on any $H_i^{(m)}$ node, and examine the linear
combination $U(1)_{Y,H}^{(m)}$ as in equation (\ref{eqn:U1yh and
  U1myh}).  From (\ref{eqn:chiral masslessness constraint}) the
conditions on a node $H_i^{(m)}$ necessary for $U(1)_{Y,H}^{(m)}$ to
remain massless are
\begin{equation}
\label{eqn:chiral excess ref}
q^{(m)}\sum_\alpha N_\alpha\,\, [\#(\ov V,\ov \alpha) - \#(\ov V,\alpha)] = 0
\end{equation}
where the sum is over hidden sector nodes in cluster $m$ and we remind the
reader that messengers are chiral under $U(1)_V$. This is equivalent to
the condition on $H_i^{(m)}$ nodes necessary for a massless $U(1)_{Y,H}$, and therefore
they are satisfied since $U(1)_{Y,H}$ is massless.
Similar statements apply for $H_I^{(m)}$ nodes. The only condition left to
satisfy is the condition on the $V$ node, given by
\begin{equation}
q^{(m)}\sum_i N_i\,\, [\#(\ov V,\ov i) - \#(\ov V,i)] = 0.
\end{equation}
This is stronger than the $V$ node condition for a necessary
$U(1)_{Y,H}$, but it is satisfied since we are considering the case
where there is no chiral excess of messengers on any $H_i^{(m)}$ node, so
that each term in square brackets is zero. Therefore
$U(1)_{Y,H}^{(m)}$ is also massless and there is yet another light $Z'$. 
If there are many such clusters, there can be many light $Z'$ bosons.

Let us consider the other case, where there is a chiral excess of
messengers on some node $H_i^{(m)}$. From \eqref{eqn:chiral excess
  ref} it is clear there is not necessarily a light $Z'$ corresponding to
$U(1)_{Y,H}^{(m)}$. However, the chiral excess induces
$U(1)_iYY$ mixed abelian anomaly since the messengers ending on
node $i$ also carry hypercharge. Such an anomaly is canceled by
Chern-Simons terms of the form $\int d^4x \, B_i \wedge F_i$ and $\int
d^4x \, \phi_i \, F_Y \wedge F_Y$ where $F_i$ and $F_Y$ are the field
strengths of $U(1)_i$ and $U(1)_Y$ and $B_i$ and $\phi_i$ are a two-form
and its Hodge dual zero-form. The Chern-Simons terms introduced to cancel anomalies
of this type can play an important role in dark matter annihilation, as
we will now discuss.

\subsection{Dark Matter and a Possible Monochromatic  $\gamma$-ray Line}
\label{subsec:dark matter}

It has been known for many years that string consistency often
requires the presence of hidden sectors which can give rise to
interesting dark matter candidates. In the $E_8\times E_8$ heterotic
string, the standard model spectrum is typically constructed from one
of the two $E_8$ factors. The other factor generically gives rise to
another gauge sector which interacts with the visible sector only
gravitationally. In weakly coupled type II orientifold
compactifications and F-theory, ``filler branes'' which do not
intersect the standard model branes are often required for tadpole
cancellation. See, for example, \cite{Cvetic:2001nr}. These interact
gravitationally with the standard mode, but not via gauge
interactions.

It is also possible that nature contains a dark matter sector which
couples weakly to the standard model, but nevertheless can
exhibit dark matter annihilation into standard model particles via
gauge interactions or suppressed couplings to visible sector
particles.  In the last six months many models of this type have been
explored, due in part to the possible experimental observation of a
$\gamma$-ray line from dark matter annihilation near the galactic
center \cite{Weniger:2012tx,Tempel:2012ey,Boyarsky:2012ca,Su:2012ft}. Regardless of whether this signal
survives further scrutiny, particularly by the Fermi LAT collaboration
itself, it is important to discuss whether dark matter candidates in our
theories can annihilate via processes with visible sector particles in the final
state, particularly photons.

\subsubsection{Annihilation via Axionic Couplings and $Z'Z\gamma$ Vertices}
\label{sec:dark matter annhilation}
We showed in section \ref{subsec:z' physics} that the stringy hidden valleys
we study generically have a rich structure of $U(1)$ physics and axionic
couplings. These can have important consequences for dark matter annihilation.

Let us briefly review two ideas in the literature which are very
common in our models and give rise to dark matter annihilation processes
with photons in the final state. The first utilizes an intermediate
anomalous $Z'$ boson to give the dark matter annihilation process
$\chi \ov \chi \rightarrow Z' \rightarrow Z \gamma$.  This was
proposed a few years ago in \cite{Dudas:2009uq} and more recently in
\cite{Dudas:2012pb} after the possible observation of the
$\gamma$-line. The key feature is an anomalous $U(1)$ symmetry under
which dark matter is charged. Anomaly cancellation via the
Green-Schwarz mechanism requires the presence of axionic couplings
which give an effective $Z' Z \gamma$ vertex that makes the
annihilation process possible.  
One difficulty is that the annihilation cross section is
suppressed by the $Z'$ mass, which is typically very large.
See section \ref{sec:z' mass}.

This possibility is extremely common in our models. Structurally, all
that is needed is dark matter charged under some symmetry $U(1)_X$ and a
$U(1)_XYY$ anomaly. In our models there are many $U(1)$ symmetries
which may play this role and this possibility could be checked on a quiver
by quiver basis. However, $U(1)_V$ is a distinguished $U(1)$
symmetry in \emph{all} of our quivers. As we have argued in section
\ref{subsec:z' physics}, hidden hypercharge quivers always have messengers which
contribute to the $U(1)_VYY$ anomaly coefficient and the quiver
exhibit a $U(1)_VYY$ anomaly unless the contribution from the visible
sector precisely cancels those of the messengers. We have argued that
this never happens for extensions of three-node MSSM quivers, and
therefore a $U(1)_VYY$ anomaly is generic in those models.  In
addition, even if the hidden sector is not hypercharged there is
almost always a $U(1)_VYY$ anomaly just from the visible sector
contribution. Thus, dark matter charged under $U(1)_V$ can nearly
always realize the scenario of \cite{Dudas:2009uq}, at least structurally. By the definition
of $U(1)_V$, such dark matter is messenger dark matter, which we will
discuss.

Another possibility was recently proposed \cite{Fan:2012gr} which
utilized similar axionic couplings.  The theory has a hidden sector
with a non-anomalous $U(1)_X$ and an $SU(N)$ gauge factor with quarks
carrying appropriate $U(1)_X$ charge to give rise to neutral or
$U(1)_X$-charged hidden sector pions. There are axionic couplings of
the form $\phi F_Y \wedge F_Y$ and $\phi G \wedge G$ where $F_Y$ and
$G$ are the hypercharge and $SU(N)$ field strengths, respectively. The
$U(1)_X$-charged pions are stable due to being the lightest $U(1)_X$
charged particles and are identified as dark matter. They can
annihilate to $U(1)_X$-neutral hidden sector pions which can then
decay to photons via the axionic couplings. 
See section \ref{sec:fan-reece example} for a concrete realization similar to this possibility in
a stringy hidden valley.
 
Our models frequently realize axionic couplings similar to these. In
certain cases it is possible to add these axionic couplings by hand,
as in \cite{Fan:2012gr}. The more interesting case, however, is when
they are required for anomaly cancellation. As argued in section
\ref{subsec:z' physics}, there is a $U(1)_VU(1)_i^2$ anomaly for any
$H_i$ node and also a $U(1)_VSU(N_i)^2$ if $H_i$ is non-abelian,
requiring the presence of couplings $\phi_V F_i \wedge F_i$ and $\phi_V G_i
\wedge G_i$. The key coupling allowing
annihilation to photons is the axionic coupling to the hypercharge
field strength, here $\phi_V F_Y \wedge F_Y$. This is necessary for the
cancellation of a $U(1)_V Y Y$ anomaly, which nearly always exists.
Therefore our models typically have the couplings utilized in
necessary to explain dark matter annihilation via the mechanism of \cite{Fan:2012gr},
or a similar mechanism. In a given quiver, there may be anomalous $U(1)$'s other
than $U(1)_V$ which could play this role.

\subsubsection{Messenger Dark Matter and $U(1)_VYY$ Anomalies}
\label{sec:messenger dark matter}
Since all quivers we study have messenger fields to hidden gauge
nodes, one simple possibility is that dark matter is comprised of
messengers fields $M$ and $\tilde M$. Since they are quasichiral, the
messenger mass is always protected by symmetry and can therefore be
light, perhaps $\cO(GeV)$ or $\cO(TeV)$.  We see from table
\ref{table:operators-suppressions} that any perturbative
superpotential coupling of messengers to a standard model field is
string suppressed, and that similar couplings obtained via instanton
effects or couplings to singlets are also very suppressed. 
Messenger dark matter in stringy hidden valleys will
always be non-baryonic, since string consistency does not require
the addition of messengers charged under $SU(3)_{QCD}$ when extending
MSSM quivers.

Let us discuss possibilities under which messenger dark matter is
stable against decay. A simple possibility is that a symmetry ensures
stability, which is certainly possible if there is a natural symmetry
under which only messengers are charged. As shown in section
\ref{subsec:z' physics}, quivers with a hypercharged stringy hidden
sector very frequently\footnote{Always, for extensions of three-node
  quivers.}  have a massless $U(1)_{Y,H}$ which charges only the
messengers and could protect messenger dark matter candidates from
decay. In addition, any hidden hypercharge quiver and many others will
have $U(1)_F^{(m)}$ symmetries, perhaps anomalous, which charge only
the messengers to the $m^{\text{th}}$ hidden cluster.  In concrete
quivers, there could be other massless $U(1)$ symmetries which charge
the messengers, or massive $U(1)$ symmetries. Therefore, symmetries
which could protect messenger dark matter from decay are very common.

Let us discuss possible annihilation processes for messenger dark
matter in generality.  $U(1)_V$ always charges both the messengers and
some set of standard model fields, allowing for dark matter
annihilation via $\chi \ov \chi \rightarrow Z'_V \rightarrow f\ov f$
for standard model fermions $f$. In addition, unless visible sector
contributions to the $U(1)_VYY$ anomaly coefficient exactly cancel the
messenger contributions, dark matter can annihilate to photons via
$\chi \ov \chi \rightarrow Z'_V \rightarrow Z\gamma$ as discussed in
section \ref{sec:dark matter annhilation}.  However, $Z'_V$ is heavy
and dark matter annihilation cross sections are suppressed. Purely in
a low energy effective theory, though, one can treat the mass of
$Z'_V$ as a parameter and constrain the phenomenologically allowed
parameter space, as in \cite{Dudas:2012pb}. See section \ref{sec:z'
  mass} for a discussion of anomalous $Z'$ masses. In addition, any
stringy hidden valley necessarily gives rise to couplings
$\phi_V F_i \wedge F_i$ and $\phi_V G_i \wedge G_i$. Since messengers
end on $H_i$ nodes, the axionic couplings could give rise to dark
matter annihilation processes with photons in the final state, similar
to \cite{Fan:2012gr}.

Let us discuss more specific possibilities which depend on the visible
sector hypercharge embedding.  For messenger dark matter to have any
hope of being realistic in an extension of the Madrid embedding, it
must be a messenger to a cluster with $q^{(m)} = \frac{1}{2}$, which
is required for the $SU(2)_L$ charged messenger to have an
electrically neutral component $\chi$. Such a particle is a natural
WIMP candidate.  For the Madrid embedding, $U(1)_V=U(1)_b$, and dark
matter can annihilate into an anomalous $Z'_b$. Since messengers are
doublets of $SU(2)_L$, annihilation to $f\ov f$ via the process
$\chi \ov \chi \rightarrow Z \rightarrow f \ov f$ will dominate over
the process involving an intermediate $Z'_V$.  In an extension of the
non-Madrid embedding, messenger fields must end on a cluster with
$q^{(m)} = 0$ for field to have an electrically neutral component
$\chi$ and $U(1)_V=U(1)_c$. Dark matter can annihilate to $f \ov f$
via an intermediate anomalous $Z'_c$. The messengers do
not carry hypercharge, but in the case where the standard model fields
generate a $U(1)_cYY$ anomaly, dark matter can nevertheless decay as
$\chi \ov \chi \rightarrow Z'_c \rightarrow Z \gamma$. This is possible
for any extension of a three-node quiver, since there is always a $U(1)_cYY$
anomaly, as argued in section \ref{subsec:z' physics}.

\subsubsection{Hidden Sector Dark Matter}
\label{sec:hidden dark matter}
Another possibility is that dark
matter is comprised of fields transforming only under hidden sector
nodes. As such, they necessarily standard model singlets.  Since
hidden sector fields are much less constrained than messenger fields,
there are more possibilities and we will therefore be brief.
Symmetries ensuring stability are similar to the messenger dark matter case, except
that hidden sector dark matter is not charged under $U(1)_{Y,H}$, the distinguished 
massless $U(1)$ common in hidden hypercharge quivers.

Since hidden sector dark matter does not carry $U(1)_V$ charge, it
cannot decay via a $Z'_V Z \gamma$ vertex. However, as argued in
section \ref{subsec:z' physics} there are broad classes of quivers
which exhibit a $U(1)_iYY$ anomaly, which introduces a $Z'_i Z \gamma$
vertex into the theory, allowing for dark matter annihilation into
photons via $Z_i'$. In such a case dark matter is necessarily charged
under $U(N_i)$ and could annihilate to photons via the axionic
couplings $\phi_V F_i \wedge F_i$ and $\phi_V G_i \wedge G_i$ as
suggested in \cite{Fan:2012gr}. This mechanism does not rely on the
propagation of a heavy $Z'_i$.  Finally, there are never $U(1)_IYY$
anomalies, since this would require hidden sector fields which carry
hypercharge. Therefore the $Z_I'Z\gamma$ vertex is not required to
exist in the low energy theory and it is unlikely\footnote{In the
  absence of $B_I \wedge F_I$ couplings there could be $\phi_I F_Y
  \wedge F_Y$ and no $U(1)_IYY$ anomaly. } that hidden sector dark
matter ending only on $H_I$ nodes will decay into photons.

\subsection{Spontaneous Global Supersymmetry Breaking}
\label{subsec:susy breaking}

In a globally consistent string
compactification, the proper framework for discussing supersymmetry
and its breaking is $\cN=1$ supergravity, where the dynamics and
stabilization of closed string moduli play an important role in
determining possible supersymmetry breaking and mediation scenarios. As
discussed, string consistency often requires the presence strongly
coupled gauge sectors which interact only gravitationally with the
standard model. It is possible that supersymmetry is broken in this
sector and gravity mediation ensues.  Such analyses require the
specification of a global string compactification with moduli
stabilized and is outside the realm of the quiver gauge theories we
study. However, in the $M_{pl} \rightarrow \infty$ limit it is natural
to study the possibility of global supersymmetry breaking. Though an embedding into supergravity may spoil\footnote{For example, in a string compactification
the Fayet-Iliopoulos term $\xi$ depends on closed string moduli and may dynamically relax to zero, restoring
supersymmetry in the Fayet models we will discuss. Realizing this model in supergravity would require
stabilization at a point in moduli space with non-zero $\xi$.} the global supersymmetry 
conclusions gained via studying a quiver gauge theory, this is the best
one can do at the quiver level and the conclusions may nevertheless
hold in supergravity embeddings. In this section we will discussed  global supersymmetry
breaking scenarios in stringy hidden valleys.

One way to break supersymmetry is to embed a non-abelian gauge theory
into the low energy spectrum which exhibits strong gauge dynamics that
break supersymmetry \cite{Affleck:1984xz}.  A prototype which has been
studied extensively is
$\cN=1$ supersymmetric QCD with $SU(N_c)$ gauge
symmetry and $N_f$ vector-like flavors \cite{Affleck:1983mk}.  Metastable supersymmetry
breaking \cite{Intriligator:2006dd,Intriligator:2007py} is a common and intriguing
possibility, in SQCD and in general.   In addition, classic supersymmetry breaking models which
do not utilize strong gauge dynamics have been realized in simple
D-brane quivers \cite{Florea:2006si,Aharony:2007db}, where D-instantons
play a crucial role in determining scales in the model. Global realizations include \cite{Cvetic:2007qj,Cvetic:2008mh}.  We find that supersymmetry breaking via SQCD
and a retrofitted Fayet model similar to those of
\cite{Aharony:2007db} can appear naturally in the models we study.

One important feature that we must consider with either SQCD or Fayet breaking is that
messenger fields often play a crucial role. In such a case supersymmetry breaking can
give vacuum expectation values to the scalar components of the messengers, breaking the
MSSM gauge group in the common case of non-singlet messengers. In particular, in
extensions of the Madrid embedding the messengers carry $SU(2)_L$ charge and supersymmetry
breaking involving messengers VEVs would trigger electroweak symmetry breaking. 
For simplicity we will avoid this possibility,
when necessary, in the examples of section \ref{sec:examples}. 

\subsubsection{Breaking Supersymmetry with SQCD}
Since we take hidden sector gauge group $G_H
= \prod_i U(N_i)$, realizations of supersymmetry breaking with strong
gauge dynamics necessarily require an $U(N_i)$ gauge group. In a
generic hidden sector there could be many such factors with rich gauge
dynamics, but for simplicity we will restrict our attention to the
possibility of a single non-abelian factor with gauge group $U(N_c)$
with $N_f$ flavors which are vector-like with respect to $U(N_c)$. All
flavors are necessarily bifundamentals, and for simplicity we also require that they
have one end on a common node which is not the $U(N_c)$ node.  Given
these restrictions, it is natural to classify the possibilities according
to whether the flavors are messengers or hidden sector fields. We refer to these scenarios
as ``messenger SQCD'' and ``hidden sector SQCD'', respectively. Of
course, hybrid scenarios are also possible if $U(N_c)$ is an $H_i$
node.

Over time it has been shown that SQCD can break supersymmetry for many
values of $N_f$ and $N_c$, originally in the confined $N_f < N_c$ regime
in \cite{Affleck:1983mk}. More recently it has been shown \cite{Intriligator:2006dd,Intriligator:2007py}
that SQCD can give rise to metastable supersymmetry breaking in the free magnetic
range $N_c + 1 \le N_f < \frac{3}{2}N_c$.
For a recent discussion of these ideas and their history, see \cite{Intriligator:2007cp}.

\vspace{.6cm}
\noindent \emph{Messenger Flavors} 
\vspace{.2cm}

\begin{figure}[ht]
 \centering
 \includegraphics[scale=0.9]{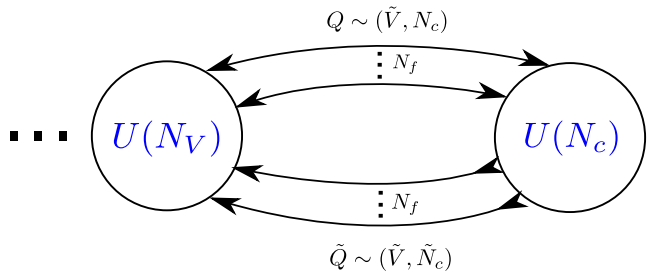}
  \caption{SQCD with messenger flavors. In each node the gauge group is showed in blue. $U(N_V)$ is the visible
sector node, and $U(N_c)$ is an $H_i$ type hidden sector node. The messenger sector is made of $N_f$ copies of $Q$ and $\tilde Q$.}
  \label{fig:sqcd}
\end{figure}

If the $U(N_c)$ node is an $H_i$ node, the SQCD flavors can end
on a visible sector node $V$ with $U(N_V)$ gauge symmetry and $N_V\in\{1,2\}$. The
flavors are what we have been calling ``messenger'' fields, where this
should not necessarily be confused with messengers of gauge mediated
supersymmetry breaking. The quiver takes the form shown in figure \ref{fig:sqcd}
and the field content beyond the standard model is
$N_f$ copies of $Q\sim(\ov V,N_c)$ and $\tilde Q \sim (\ov V, \ov
N_c)$, and the flavors are chiral with respect to the trace
$U(1)_V$ of $U(N_V)$. To avoid
detailed analyses of supersymmetry breaking scenarios for different
values of $N_f$ and $N_c$, we will utilize facts about $SU(N_c)$ SQCD
despite the fact that our gauge group is $U(N_c)$. This is certainly a
valid assumption at scales below the mass of the non-anomalous $Z'$
boson associated to the trace $U(1)$ of $U(N_c)$.  We will give a
concrete example of these models in section \ref{sec:non-madrid extensions}. Let us discuss
some generic features here. 

An important feature of these realizations of SQCD is that the
mass term $Q\tilde Q$ is protected by symmetry but can be generated at
a low scale via D-brane instantons or couplings to
singlets. In the absence of this symmetry, the flavors will typically
obtain a large mass far above the confinement scale $\Lambda_{N_c}$,
giving a pure SQCD theory at low energies which does not break
supersymmetry.  We view this as an advantage of these models and assume
that the masses of the flavors is far below the confinement scale. A
natural concern in this theory is that it may be difficult to realize
the Affleck-Dine-Seiberg non-perturbative superpotential which plays
an important role in supersymmetry breaking, since $Q\tilde Q$
explicitly appears and is forbidden by symmetry. However, it is known
\cite{Haack:2006cy} that a gauge invariant\footnote{In these
  constructions the non-gauge invariance of $Q\tilde Q$ is compensated
  for by the non-gauge invariance of a closed-string modulus appearing
  in the correction.} ADS superpotential can be generated even in the
case where $Q\tilde Q$ carries net anomalous $U(1)_V$ charge. 
We have argued in section \ref{subsec:z' physics} that $U(1)_V$
is always anomalous in models with stringy hidden sectors.
Given these arguments, one can apply standard techniques of 
supersymmetry breaking via SQCD with various various of $N_f$ and $N_c$.

For SQCD with messenger flavors in our models, the allowed values of
$N_f$ and $N_c$ are constrained by the fact that messengers are added
to cancel some non-zero T-charge, and the T-charges are concretely
determined by possible visible sector realizations of the MSSM. For
example, in extensions of the three-node Madrid hypercharge embedding
the only possible non-zero T-charge is $T_b = \pm 2n$ for $n \in
\{0,\dots,7\}$, as discussed in section \ref{sec:three node MSSM},
which constrains the allowed values of $N_f$ and $N_c$ via the
equation
\begin{equation}
  2N_fN_c = 2n.
\end{equation}
We have assumed a single SQCD node of $H_i$ type. Due to the $\text{mod}\, 3$
condition for $U(1)$ nodes in equation (\ref{eqn:chiral tadpole constraint}),
SQCD extensions of the non-Madrid embedding must have $N_c$ which is a not a multiple
of $3$, as must any stringy hidden sector with an SQCD node attached to a visible
sector $U(1)$ node.
It is also possible to write down the allowed values of $N_f$
and $N_c$ for extensions of higher-node MSSM quivers. 
There are allowed values of $N_f$ and $N_c$ which break supersymmetry via the ADS
superpotential.

Finally, for SQCD supersymmetry breaking with messenger flavors it is possible that
the messengers fill out non-trivial standard model representations, in which case the
ADS superpotential Higgses $G_{MSSM}$. The only possibility for the messenger flavors to
be standard model singlets is an in extension of the non-Madrid embedding with $q^{(m)}=0$.
See section \ref{sec:non-madrid extensions} for an example.

\vspace{.5cm}
\noindent \emph{Hidden Sector Flavors}
\vspace{.2cm}

The other possibility is that the $U(N_c)$ gauge theory which breaks supersymmetry
is realized on an $H_I$ type node, in which case the $N_f$ flavors cannot be messenger fields.
In this case there is no constraint on the allowed values of $N_f$ and $N_c$ since the flavors
are hidden and they are not required to cancel a T-charge.
Hidden sector fields are not required to be quasichiral and therefore
in this case it is possible to realize vanilla SQCD with vector-like flavors. However,
such flavors do not have masses protected by symmetry and are very heavy at a generic point
in the moduli space of a string compactification. If so, the flavors can be integrated out,
giving pure glue SQCD at low energies which does not break supersymmetry.

In clusters with $q^{(m)}=0$, the hidden sector SQCD flavors could also be quasichiral
bifundamentals with protected masses, giving rise to a scenario very
similar to that of the messenger flavor case. However, compared to the
messenger flavor case the structure superpotential couplings is
different, according to table \ref{table:operators-suppressions},
and the possibilities are not as constrained.

\subsubsection{Breaking Supersymmetry via a Retrofitted Fayet Model}
\label{sec:fayet}

In \cite{Aharony:2007db} a retrofitted Fayet model which broke
supersymmetry was presented in a simple quiver. We remind the reader
that a Fayet model generically contains a $U(1)$ symmetry with a
non-zero Fayet-Iliopoulos term $\xi$ and some number of fields charged
under the $U(1)$. Since the F-term and D-term equations cannot be
simultaneously satisfied, supersymmetry is broken. Given the many
$U(1)$ symmetries in our hidden sectors, it seems natural that this
model of supersymmetry breaking could be realized.

We would like to realize the Fayet model without needing to specify a
concrete spectrum or hypercharge embedding. There are typically many
heavy anomalous $Z'$ bosons in a given quiver, but as emphasized in
\cite{Intriligator:2005aw} the corresponding D-term equations should
not be imposed since the $Z'$ bosons can be integrated out of the low
energy theory. Therefore, successful Fayet models should utilize
massless $U(1)$ symmetries. Fortunately, in hidden hypercharge quivers there is
typically a light $Z'$ corresponding to the gauge symmetry
$U(1)_{Y,H}$, as discussed in section \ref{subsec:z' physics}. We will
study the possibility of a single cluster hidden sector, though the
arguments we present can be trivially generalized to the case of
multiple cluster hidden sectors.  Given that the hypercharged
stringy hidden sector has a single sector, we will rescale $U(1)_{Y,H}$ by $1/q^{(m)}$
to give a symmetry $U(1)_F$ for simplicity. This allows the discussion to
proceed without reference to the value of $q^{(m)}$.

Let us discuss how $U(1)_F$ can play a role in supersymmetry breaking.
Recall from section \ref{sec:hidden sectors sec 2} that it is 
\begin{equation}
U(1)_F = \sum_\alpha U(1)_\alpha
\end{equation}
where the sum is over all hidden sector nodes. The only fields charged
under this symmetry are the messengers $\Phi_+^I$ and $\Phi_-^J$,
which carry positive and negative $U(1)_F$ charge, respectively. We assume that
the messengers have non-zero mass. From table
\ref{table:operators-suppressions}, we see that couplings of type
$\OM{i}\OH{j}$ are forbidden in perturbation theory, while couplings
of type $\cO_V \cO_M$ and $\cO_V \cO_M \cO_H$ are at least string
suppressed if the visible sector contains the exact MSSM
spectrum. These couplings are either absent or generated via
instantons or couplings to singlets. In the case that they are
present, each coupling has an additional prefactor, given for example
by worldsheet instantons\footnote{ \label{foot:WS-instanton}We are
  careful here to utilize the worldsheet instanton prefactor as the
  small parameter, rather than the D-instanton prefactor. This is
  because the same D-instanton which generates the messenger mass
  $\cO_{\text{mess}}$ will also generate couplings of the form
  $\cO_{\text{mess}}\cO_H$ for perturbative couplings $\cO_H$. In this
  case the D-instanton suppression of the operators cannot be tuned
  independently, whereas it is possible that the worldsheet instanton
  prefactors of each coupling could be tuned independently. The validity
  of our results in a given setup depend, of course, on whether a
  small parameter $\eps$ exists. } in type IIa, which in principle
allows them to be very small. Call $\eps$ the parameter of the largest
such coupling.

In the limit where couplings of messengers to visible
and/or hidden sector fields are absent, or in the limit
$\eps\rightarrow 0$, the superpotential takes the form
\begin{equation}
W = W_{MSSM} + m_{IJ} \,\, \Phi^I_+ \Phi_-^J,
\end{equation} 
where the mass matrix is generated via instantons or couplings to singlets.
The relevant F-term and D-term equations are
\begin{align}
  F_{\Phi^I_+} &= m_{IJ}\, \Phi^J_- =0 \qquad \qquad F_{\Phi^J_-} = m_{IJ}\, \Phi^I_+ =0 \nonumber \\ \nonumber \\ &D_F = \xi + g^2 (\sum_I |\Phi_+^I|^2 - \sum_I |\Phi_-^I|^2) = 0.
\end{align}
Supersymmetry is broken since these conditions cannot be simultaneously satisfied for non-zero $\xi$.

It is possible that the theory exhibits has an
R-symmetry, with the requirements $R_+ + R_-=2$ for the R-charges of
the positively and negatively charged fields. If the R-symmetry
exists, taking $\eps$ small but finite gives a small explicit
R-breaking which modifies the F-terms in a way that will typically
restore supersymmetry. The non-supersymmetric vacuum is metastable and
is separated from the supersymmetric vacuum by a distance in field space which
is proportional to an inverse power of $\eps$.  These are essentially the
arguments for metastability presented in \cite{Intriligator:2007py}, where a small
explicit R-symmetry breaking parameter $\eps$ sets the distance to the
supersymmetric vacuum, and thus the lifetime of the metastable state.

Let us discuss one caveat. As specified currently, we have utilized
the massless $U(1)_F$ symmetry that exists for any quiver with
$q^{(m)}\ne 0$ to realize a Fayet model of supersymmetry breaking,
where the messenger fields play the crucial role in the model and must
obtain vacuum expectation values. Since $q^{(m)}\ne 0$, the messengers
necessarily carry hypercharge and their vacuum expectation value
spontaneously breaks $U(1)_Y$. Additionally, in extensions of the
Madrid embedding the messengers are doublets of $SU(2)_L$ and
therefore spontaneously break it. There is a simple way to avoid
spontaneous breaking of the standard model without losing the Fayet
model, however. Construct a Fayet model as specified in an extension
where the messengers are $SU(2)_L$ singlets, such as in an extension
of a non-Madrid quiver. Keep the same spectrum, but set $q^{(m)}=0$ so
that the messengers do not have hypercharge. Nevertheless, $U(1)_F$ is
a good massless symmetry and the breaking of supersymmetry breaking is
as above, but without breaking $G_{MSSM}$. Thus, we can avoid this issue in
a subclass of models with $q^{(m)}=0$.  

We emphasize that this is only one possibility of breaking
supersymmetry, though it is common.  In concrete quivers there
may be additional D-terms corresponding to other massless $U(1)$
symmetries that one should impose, giving more complicated
realizations of the Fayet model.  There may also be completely
different methods of breaking supersymmetry, in particular there are
likely realizations of the O'Raifeartaigh and Polonyi quivers of
\cite{Aharony:2007db}. Our main point in this section is simply that
$U(1)_F$ exists in broad classes of quivers and can give a realization
of the Fayet model.  We will present a concrete example in
section \ref{sec:non-madrid extensions}.

\subsubsection{Possibilities for Mediation}
In a successful supergravity embedding one could explore the possibility
of gravity mediation. Since we are studying the $M_{pl}\rightarrow \infty$
limit and a supersymmetry breaking hidden sector in our models is connected
to the visible sector, it is natural to explore alternatives. Given the importance
of messenger fields and D-instantons in our quivers, as well as the presence of heavy
and light $Z'$ gauge symmetries, SUSY breaking could naturally be communicated by a combination of gauge mediation,
$Z'$ mediation, and D-instanton mediation.
 
\vspace{.5cm}
\noindent \emph{Gauge Mediation}
\vspace{.3cm}

It is natural to consider gauge mediation \cite{Giudice:1998bp}, since
there are always messenger fields. In section \ref{subsec:susy
  breaking} we discussed three common possibilities for supersymmetry
breaking in stringy hidden valleys, which we called messenger SQCD,
hidden SQCD, and Fayet models.  In messenger SQCD and the Fayet
models, the ``messenger fields'' themselves play a role in
supersymmetry breaking, and thus should not be identified with
possible messenger fields used for gauge mediation\footnote{Remember
  that a messenger field, for us, is a field which is a bifundamental
  of a visible sector and hidden sector node, regardless of
  interpretation in any concrete physical scenario.}. However, hidden
SQCD can break supersymmetry without utilizing the messenger fields.
In hidden SQCD or any other scenario which breaks supersymmetry
without utilizing messenger fields, it is possible to consider the
messenger fields as messengers of gauge mediation.

Let us begin by considering the case where messengers are only added
for the sake of string consistency. In this case, messengers are
always trivial under $SU(3)_{QCD}$ and therefore gluino masses are not
generated at one loop. Messenger masses are protected by $U(1)_V$ symmetry and
therefore have good reason to be light. This is in contrast to other models in the
literature, where messengers are vector like with respect to all symmetries and 
therefore light messengers require a considerable fine-tuning.

In extensions of the Madrid embedding,
messengers are charged under $SU(2)_L$ and will generate soft masses
for the corresponding gauginos. If the supersymmetry breaking hidden
sector is hypercharged, the messengers carry hypercharge and
will generate a soft mass for the bino. In extensions of the
non-Madrid embedding, messengers are singlets of $SU(3)_{QCD}\times
SU(2)_L$ but carry hypercharge if the SUSY breaking hidden sector is
hypercharged, generating a bino soft mass. Without adding
non-required messengers, other mechanisms must account for the soft
masses of the gluinos and squarks.

If one is willing to abandon our guiding principle of only adding fields for the sake of string
consistency, it is possible to add messengers transforming as $3+\ov 3$ of
$SU(3)_{QCD}$, which can be realized as $(a, i)$ + $(\ov a, \ov i)$. If the
$H_i$ node is a $U(1)$, they could also be chiral under $U(1)_i$. In
extensions of the Madrid embedding, the messengers added for
string consistency are $(1,2)_Y + (1,2)_{-Y}$, but can never fill out a
$5+\ov 5$ of $SU(5)$ with the added $3+\ov 3$, since they are chiral
under $U(1)_b$. The messenger masses of the doublets and triplets are generically
different, and the gluino will be lighter than the bino and wino
if the $3 + \ov 3$ are vector like under $U(1)_i$. 
In extensions of the non-Madrid
embedding, it is possible to add both doublet and triplet messengers, possibly
filling out a $5 + \ov 5$. If the messengers required for consistency carry
hypercharge there will be additional contributions to the bino mass beyond
those of minimal gauge mediation from the $5+\ov 5$. If the required
messengers do not carry hypercharge, then they do not participate in gauge mediation.
In conclusion, there is \emph{no} realization of minimal gauge mediation which
utilizes the required messengers.
 
\vspace{.5cm}
\noindent \emph{$Z'$ Mediation}
\vspace{.3cm}

Another possibility for mediation of supersymmetry breaking is via
$Z'$ gauge bosons, which has been discussed in both field theoretic \cite{Binetruy:1996uv,Dvali:1996rj,Langacker:2007ac,Langacker:2008ip}
and string theoretic models \cite{Verlinde:2007qk,Grimm:2008ed}.
In \cite{Langacker:2007ac} a class of models was proposed where the
dominant mediation mechanism
was via a $U(1)'$ gauge interaction which charged fields in both the visible sector
and SUSY breaking hidden sector. In particular, the absence
of messengers ensured the absence of leading contributions from gauge mediation. 
This is not possible in a stringy hidden valley, since messengers are generic.

It is still important to consider whether $Z'$ mediation is a viable
possibility for the partial mediation of supersymmetry breaking. For a
generic stringy hidden valley, there exists no\footnote{With the
  exception of hypercharge, which could charge the messengers.}
massless $U(1)$ which charges both the visible sector and either
messenger or hidden sector fields. In particular, though $U(1)_{Y,H}$
is massless for a wide class of quivers, it never charges the standard
model fields.  However, in any specific quiver it is possible that
there are additional massless $U(1)$'s and $Z'$ mediation could be
realized.

Since there does not generically exist a non-anomalous $U(1)$ which
charges both visible sector and hidden sector fields,
the only $U(1)$ symmetries which could give rise to $Z'$ mediation for a
generic stringy hidden valley are anomalous. For example,
$U(1)_V$ charges both standard model fields and messengers, but it is
always anomalous and $Z'_V$ has a large St\" uckelberg mass. Mediation
of supersymmetry breaking via heavy $Z'$ bosons must
be subleading unless its mass is fine tuned
to a low scale.

\vspace{.5cm}
\noindent \emph{D-Instanton Mediation}
\vspace{.3cm}

D-instantons can generate non-perturbative corrections to the
superpotential which couple visible sector fields to hidden sector
fields, i.e. $\cO_V \cO_H$ couplings. As emphasized in table
\ref{table:operators-suppressions}, these couplings are forbidden in
perturbation theory but can be generated with exponential suppression
via D-instantons. In \cite{Buican:2008qe} it was suggested to that
these non-perturbative corrections can mediate supersymmetry breaking,
giving rise to so-called D-instanton mediation. Though the original work only
considered couplings of the form $\cO_V \cO_H$, it is also possible to
consider D-instanton mediation via couplings of the form $\cO_M \cO_H$
and $\cO_V \cO_M \cO_H$ which involve the messenger fields. These
couplings are forbidden and string suppressed in perturbation theory,
respectively. In certain quivers, some of these couplings are
necessarily generated. For example, the messenger mass terms must be
obtained via D-instantons or coupling to singlets, and in the former
case an instanton which generates a messenger mass term $M\tilde M$
will also generate a coupling of the form $M \tilde M \cO_H$ for any
perturbative hidden sector coupling $\cO_H$. However, the precise structure of
the mediation is very model-dependent.

\subsection{Implications of $Z'$ Masses for Low Energy Physics}
\label{sec:z' mass}

Having discussed the importance of $Z'$ bosons for the structure of stringy
hidden valleys, including possibilities for dark matter and supersymmetry breaking,
let us briefly comment on the implications of their masses for low energy physics. 

As discussed, light $Z'$ bosons obtain a mass via the standard Higgs
mechanism, and could be $\cO(TeV)$.  Consider heavy $Z'$ bosons. If
arising from a string compactification they have a string scale St\"
uckelberg mass, which could be fine-tuned to low scales, but is
typically near the GUT or Planck scale. This is an important consideration when
discussing the low energy physics of gauge theories with anomalous $Z'$ bosons. 
If the $Z'$ considered purely in an
effective field theory where anomalous $U(1)$ symmetries require the
presence of Chern-Simons terms, the $Z'$ mass is a parameter.
Regardless of its origin, it is useful to discuss what occurs at low energies
as the $Z'$ mass parameter is gradually lowered from Planck scale to weak scale.

For example, it is interesting to consider the mass $m_V$ of the heavy $Z'_V$ gauge boson. This field always couples to
both messenger fields $\psi$ and some standard model fermions $f$. If $m_V$ is Planck scale, it serves as a barrier
between the visible and hidden sectors by suppressing the cross sections of processes such as $\psi \ov \psi \rightarrow Z'_V \rightarrow f \ov f$
or $\psi \ov \psi \rightarrow Z'_V \rightarrow Z \gamma$, making it unlikely that the dark matter scenario of \cite{Dudas:2012pb} can account for the
supposed $\gamma$-line \cite{Weniger:2012tx} coming from the center of the galaxy. As $m_V$ is lowered, so is the $Z'_V$ barrier and these annihilation processes give rise to
interesting signals.

\subsection{Generality of Results and Higher-Node Extensions}
Though some of our discussion has focused on extensions of three-node MSSM quivers, our results
are much more general. 

Generically, most of the physics we have discussed throughout this
section requires only a hidden sector with messengers carrying some
non-zero $T_V$ charge. This occurs when the MSSM fields carry some net
$T_V$ charge which must be canceled. This is true of nearly any
stringy hidden valley of the type we study, regardless of the number
of nodes in the MSSM quiver which has been extended.  The only
possible loophole is if all of the T-charge conditions
(\ref{eqn:chiral tadpole constraint}) are satisfied but the
masslessness conditions (\ref{eqn:chiral masslessness constraint}) are
not, though the conclusions may hold in this case as well.

We can be more specific. Consider the three-node Madrid embedding.
The salient feature which governs all extensions of the Madrid
embedding is that hidden sectors are connected to the visible sector
via messengers ending on the $U(2)_b$ node, since $T_b\ne 0$ for all
three node quivers with the Madrid embedding. Therefore, hidden
sectors added to three-node Madrid quivers can be added to \emph{any}
higher node quiver with the same $T_b$ charge.  This is actually a
very common possibility.  To realize the MSSM gauge group at low
energies, an arbitrary higher node MSSM quiver has gauge group
$U(3)_a\times U(2)_b\times \prod_i U(1)_{c_i}$.  Consider the
hypercharge embedding
\begin{equation}
 U(1)_Y = \frac{1}{6} U(1)_a+ \sum_i q_iU(1)_{c_i }
\end{equation}
and possible MSSM realizations. $Q$, $L$, $H_u$, and $H_d$ must
transform under the $U(2)_b$ node, and $e^c$, $u^c$ and $d^c$ must
not.  The possible $T_b$ charges for an MSSM quiver are again $T_b =
\pm 2n$ with $n\in \{0,\dots,7\}$, and therefore the hidden sectors
added to three-node Madrid quivers can be added to many higher node
quivers, including the higher node Madrid embeddings with
$q_i=\frac{1}{2} \,\,\,\forall i$.  Such a quiver may require other
additions to satisfy T-charge or M-charge conditions on other nodes,
which is not necessary in the three-node case. Similar statements can
be made about the non-Madrid embedding.

\section{Example Quivers}
\label{sec:examples}

In section \ref{sec:general physics} we discussed many aspects of
physics which are generic in stringy hidden valleys, or at least a
broad subclass, and discussed implications for low energy physics. We
discussed the implications of $U(1)_V$ charged messengers for
superpotential couplings, the ubiquity of both heavy and light $Z'$
bosons, and axionic couplings required for anomaly cancellation.
These ingredients had strong implications for both dark matter and
supersymmetry breaking in these models.

In this section, we only seek to exemplify the more generic
discussions of previous sections in concrete quivers with interesting
low energy physics. We will discuss three extensions of Madrid
quivers and two extensions of non-Madrid quivers.

\subsection{Hidden Extensions of the Madrid Embedding}
\label{sec:madrid extensions}
Let us study simple extensions of three-node MSSM
quivers in the Madrid embedding. Before
adding hidden sectors, the possible T-charges and M-charges of Madrid
quivers with the exact MSSM spectrum are given in equation
(\ref{eqn:Madrid T-charge M-charge}). Exploiting a quiver symmetry, we
can take $T_b>0$ without loss of generality\footnote{From the D-brane perspective, we can choose a brane on either $\pi_{b}$ or $\pi_b{'}$ to be
the orientifold-image brane. This redundancy gives the mentioned quiver symmetry, so that there is
a physically equivalent quiver with $T_b < 0$ for every quiver with $T_b >0$ via mapping $b \leftrightarrow \ov b$.}, so that messengers added
for consistency must end on the $U(2)_b$ node and be negatively
charged under the $U(1)_b$.

\subsubsection{Dark Matter Annihilation into Photons via an Intermediate $Z'$}
\begin{figure}[ht]
\centering
\includegraphics[scale=0.7]{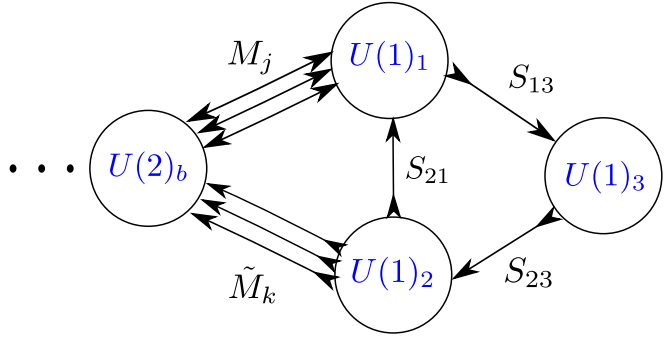}
\caption{An example quiver which allows for dark matter annihilation
  into a photon via a $Z' Z \gamma$ vertex. $U(1)_1 + U(1)_2 + U(1)_3$
  is a massless $U(1)$, giving a light $Z'$ boson.}
\label{fig:chiral_excess}
\end{figure}

In this section we will demonstrate possibilities for dark matter and light $Z'$ bosons discussed in section
\ref{sec:general physics}. 
The stringy hidden valley is
demonstrated in figure \ref{fig:chiral_excess} and has three hidden
sector nodes, two of $H_i$ type and one of $H_I$ type. 
This hidden sector has $T_b = -6$ and can therefore be added
consistently to any three-node MSSM quiver in the Madrid embedding
with $T_b=6$. Since there is only a single hidden sector, we define
$q\equiv q^{(m)} \ne 0$ and the messengers $M_j$ and $\tilde M_k$ have
hypercharge $\pm q$, respectively.  We take $q\in \bZ + \frac{1}{2}$
in order to messengers particles with fractional electric charge.
Messenger masses are protected by symmetry but can be generated
non-perturbatively by instantons or via a perturbative coupling
$S_{21}M_j\tilde M_k$ if $S_{21}$ obtains a vacuum expectation
value. There is a perturbative superpotential term
$S_{13}S_{32}S_{21}$ corresponding to the closed hidden sector loop.

We have constructed this quiver to exemplify some of the generic discussion of $Z'$
physics in section \ref{subsec:z' physics}. The linear combination\footnote{Recall $U(1)_F=U(1)_{Y,H}/q^{(m)}$.}
\begin{equation}
U(1)_F = U(1)_1 + U(1)_2 + U(1)_3
\end{equation}
is massless and gives a light $Z'$ boson, as expected for hidden
hypercharge quiver which extends a three-node MSSM quiver.  Since
there is a chiral excess of messengers on both $U(1)_1$ and $U(1)_2$,
the associated boson $Z'_1$ and $Z'_2$ obtain a St\" uckelberg mass,
as can be verified by explicitly checking equations (\ref{eqn:chiral
  masslessness constraint}) for each $U(1)$. This is in agreement with
the fact that there are $U(1)_1YY$ and $U(1)_2YY$ anomalies, due to
the messenger fields.

The hidden sector contains three singlet fields which could be dark
matter candidates. Any mass term in these fields is forbidden in
perturbation theory, but can be generated by instantons at a
suppressed scale, perhaps giving $\cO(GeV)$ or $\cO(TeV)$ dark
matter. The $U(1)_1YY$ and $U(1)_2YY$ anomalies are canceled via the
introduction of Chern-Simons terms. The necessary terms generate
vertices $Z'_1Z\gamma$ and $Z'_2Z\gamma$ in the low energy theory and
dark matter can annihilate into $Z\gamma$ via intermediate $Z'_{1,2}$
bosons, as suggested in the dark matter scenario of
\cite{Dudas:2009uq,Dudas:2012pb} to explain the tentative $\gamma$
line \cite{Weniger:2012tx}.  As emphasized in section \ref{sec:general
  physics}, though, in string compactifications these $Z'$ bosons have
a string scale St\" uckelberg mass within a few orders of $M_{pl}$, as
opposed to the weak scale $Z'$ of \cite{Dudas:2012pb}.

Dark matter realized as the neutral component of the messengers is
also a possibility in this scenario. The messengers are the only
particles charged under the non-anomalous symmetry $U(1)_F$, and
therefore are stable. It can annihilate via $Z'_{1,2}Z\gamma$ vertices
into photons, but can also annihilate via a $Z'_bZ\gamma$ vertex since
there is a $U(1)_bYY$ anomaly.  This dark matter candidate has a
protected mass and weak interactions.

In addition, axionic couplings $\phi_b F_Y \wedge F_Y$ and $\phi_b F_i
\wedge F_i$ are required for anomaly cancellation, where $F_i$ with
$i=1,2$ are the field strengths of $U(1)_1$ and $U(1)_2$. Couplings of
this form were used in a non-abelian gauge theory in \cite{Fan:2012gr}
to give dark matter annihilation processes with photons in the final state, and it would
be interesting to study their implications in this abelian model.

\subsubsection{Many Light $Z'$ Bosons}
\begin{figure}[ht]
\centering
\includegraphics[scale=0.7]{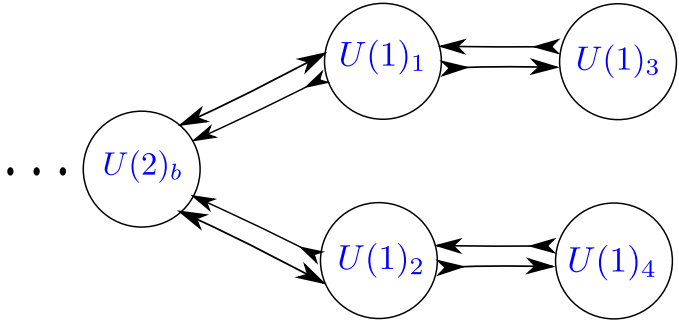}
\caption{Quiver realizing two light $Z^\prime$ bosons.}
\label{fig:manyzp}
\end{figure}

In this subsection we will exemplify the possibility of many light $Z'$ bosons as
discussed in generality section \ref{subsec:z' physics}. We study a two-cluster hidden
sector, with quiver extension given in
figure \ref{fig:manyzp} 
which can be added consistently to any three-node MSSM quiver in the
Madrid embedding with $T_b=4$.  The contribution of the hidden
clusters to the hypercharge embedding are determined by two numbers
$q^{(1)}$ and $q^{(2)}$, which must be non-zero for the messengers to
have integral electric charge. As with any hidden hypercharge quiver extending a three-node MSSM quiver,
$U(1)_{Y,H}$ satisfies the equations (\ref{eqn:chiral masslessness constraint}),
which here is given by
\begin{align}
U(1)_{Y,H} &= q^{(1)}\,\,[U(1)^{(1)}_1 + U(1)^{(1)}_2] +
q^{(2)}\,\,[U(1)^{(2)}_1 + U(1)^{(2)}_2] \nonumber \\ &\equiv U(1)_{Y,H}^{(1)} + U(1)_{Y,H}^{(2)}
\end{align}
and the boson $Z'_{Y,H}$ is light. From the general discussion of
section \ref{subsec:z' physics}, the symmetries $U(1)_{Y,H}^{(1)}$ and
$U(1)_{Y,H}^{(2)}$ are also massless, since there are two hidden
clusters with no chiral excess of messengers on $H_i$-type nodes. It can
be explicitly checked that they satisfy the equations (\ref{eqn:chiral
  masslessness constraint}). Though it is non-generic, in this example
all hidden sector $U(1)$'s are massless.

\subsubsection{Dark Matter Annihilation and Axionic Couplings}
\label{sec:fan-reece example}

\begin{figure}[ht]
 \centering
 \includegraphics[scale=0.6]{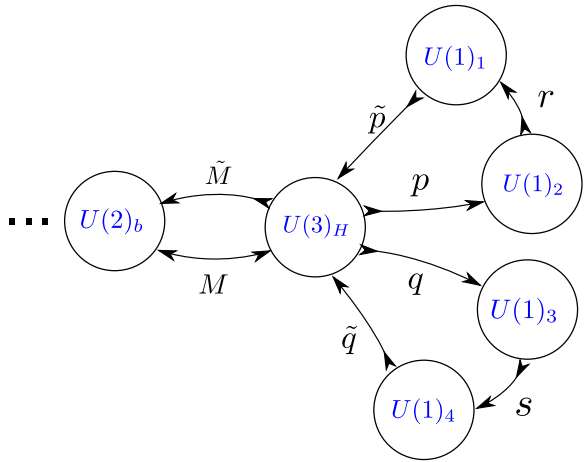} 
 \caption{A quiver similar to that of \cite{Fan:2012gr}. The quark mass terms $p\tilde p$ and $q\tilde q$ arise via couplings to singlets $r$ and $s$.}
 \label{fig:fan-reece}
\end{figure}

Here we show a simple implementation of the dark matter scenario
resembling that of \cite{Fan:2012gr}. The quiver is shown in figure
\ref{fig:fan-reece}. It contributes a $T_b$ charge of $-6$, and
therefore any quiver formed from this hidden sector and a Madrid
quiver with $T_b=6$ is consistent. In order to obtain the appropriate
$U(1)$ charges for the model, we take $U(1)_X =
U(1)_{3}+U(1)_{4}-U(1)_{1}-U(1)_{2}$ and assume that it is the
lightest of any hidden sector $Z'$s. We find this assumption plausible
since any of the individual $U(1)_I$ symmetries is anomalous.  The
quark mass terms $p\tilde p$ and $q\tilde q$ could arise via
couplings to the singlets $r$ and $s$ when those fields obtain vacuum
expectation values.

Consider the quark representations given in table
\ref{tab:fan-reece}. The dark matter candidates are the hidden sector
pions $\pi^{\pm}_H$ , $\pi^0_H$ arising from the hidden sector quarks
$p$, $\tilde p$, $q$, and $\tilde q$.  To ensure the absence of
fractionally charged massive particles, we must have $q^{(m)}\ne 0$.
There are $U(1)_b YY$ and $U(1)_b SU(3)_H^2$ anomalies and
cancellation via the Green-Schwarz mechanism requires the presence of
couplings $\phi_b F_Y \wedge F_Y$ and $\phi_b Tr(G\wedge G)$ where
$F_Y$ is the hypercharge field strength and $G$ is the field strength
of the hidden $SU(3)_H$. Hidden sector $U(1)_X$ charged pions can
annihilate to $\pi^0_H$, which can then decay into photons via the
axionic couplings.  This is very similar to the setup in
\cite{Fan:2012gr}, to which we refer the reader for more details.
One could also consider the possibility of messenger dark matter
annihilation to $\gamma \gamma$ or $Z \gamma$ via an intermediate
$Z'_b$ boson.

\begin{table}[ht]
\centering
 \begin{tabular}{c|c|c}
  Field & $U(3)_H$ & $U(1)_X$ \\ \hline
  $p$ & $\fund$ &  +1 \\
  $\tilde p$ & $\bar \fund$ &  -1 \\
  $q$ & $\fund$ &  -1 \\
  $\tilde q$ & $\bar \fund$ &  +1 \\
 \end{tabular}
 \caption{Representations of hidden sector quarks which form the hidden sector pions $\pi^\pm_H$ and $\pi^0_H$.}
\label{tab:fan-reece}
\end{table}

This quiver is a bit contrived compared to some simpler possible
realizations of \cite{Fan:2012gr}. We have two reasons for this. First, we have
intentionally added many $U(1)_I$ nodes to forbid perturbative mass
terms for hidden sector quarks, which would be string scale at a
generic point in the moduli space of a string compactification. In
this case the quark masses are typically high above the confinement
scale $\Lambda_H$ and hidden sector pions do not form.  Second, we
have chosen the quiver such that the axionic couplings which give rise
to $\pi_H^0$ decays are required, rather than put in by hand. This is
due to the structure of $SU(2)_L$ charged messengers which also
transform under $U(3)_H$, but are distinct from the hidden sector
pions. Simpler realizations may also exist.
 
\subsection{Hidden Extensions of the Non-Madrid Embedding}
\label{sec:non-madrid extensions}
Let us turn to stringy hidden valleys which extend three-node MSSM quivers with
the non-Madrid embedding. Before adding hidden sectors, the possible T-charges and
M-charges of non-Madrid quivers with the exact MSSM spectrum are given
in equation (\ref{eqn:non-Madrid T-charge M-charge}). Here messenger
fields added for consistency must end on the $U(1)_c$ node.

Extensions of the non-Madrid embedding allow for a wider variety of
possibilities than Madrid extensions. Let us state some of the
differences. Since the messengers are not charged under $SU(2)_L$, it
is possible to avoid fractionally charged massive particles by having
a hidden sector which is not hypercharged. In such a case the
messengers are MSSM singlets and any realization of supersymmetry
breaking involving messenger VEVs doesn't automatically break
$G_\text{MSSM}$.  Hidden sectors which are not hypercharged
(i.e. $q_i=q_I=0$ for all $i,I$) are more natural in non-Madrid
extensions, as discussed in section \ref{sec:fchamp}. In such a case
symmetry and antisymmetry tensor representations can give rise to
hidden sector fields, and all four bifundamental representations are
allowed, instead of just two.

\subsubsection{Breaking SUSY with a Fayet Model}

\begin{figure}[t]
\centering
\includegraphics[scale=0.5]{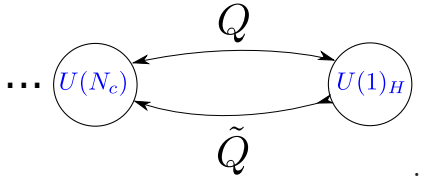}. 
\caption{A simple single-node hidden
  sector which can realize the Fayet model of supersymmetry breaking. The mass term $Q\tilde Q$
is protected by symmetry but can be generated non-perturbatively.}
\label{fig:fayet quiver}
\end{figure}

Let us consider the simplest possible abelian
hidden
sector as an example, consisting of one node $U(1)_H$. Since we are
considering extensions of MSSM quivers with $T_c>0$, we can add $N_f$
messenger fields of both types $(\ov c,H)$ and $(\ov c,\bar H)$ that
we will call $Q_i$ and $\tilde{Q}_i$, respectively. See figure \ref{fig:fayet quiver}. Together these
messenger have $T_c=-N_f$ and therefore can be added to MSSM quivers
with $T_c = N_f$ mod 3. For simplicity, we take
$N_f=2$. We can take $U(1)_{Y,H}$ to be trivial without introducing
fractionally charged massive particles, so that
\begin{equation}
U(1)_Y = -\frac{1}{3} U(1)_a - \frac{1}{2} U(1)_b.
\end{equation} 
Thus, the MSSM spectrum is augmented only by a pair of MSSM singlets
 $Q$ and $\tilde Q$.  It is easy to see that $U(1)_H$
meets the conditions (\ref{eqn:chiral masslessness constraint})
and therefore gives a light $Z'$ boson coupling only to $Q$ and
$\tilde Q$. In this quiver $U(1)_H$ is the more generic $U(1)_F$ described
in section \ref{sec:fayet}, and it is therefore natural to expect that this realizes
the Fayet model.

Let's discuss the superpotential.
As usual, the messenger mass term $Q\tilde Q$ is forbidden by the symmetry $U(1)_V$, which
here is $U(1)_c$.
Phenomenology requires that some mechanism, such as a D-instanton or couplings to
singlets with VEV, gives rise to a mass term $m\, Q\tilde Q$.
The superpotential has the generic form
\begin{equation}
W = \cO_V + \eps\,\, \cO_V \, Q \tilde Q + m Q\tilde Q
\end{equation}
where a sum over all couplings of type $\cO_V$ is implied and we have considered 
only messenger sector couplings of the form $Q\tilde Q$ and not the irrelevant
couplings $(Q \tilde Q)^k$ for $k>2$.
As discussed in section \ref{sec:fayet}, we assume that couplings of the form $\cO_V \cO_M$
can be made parametrically small by tuning a parameter $\eps$.
Then in the $\eps \rightarrow 0$ limit the superpotential is
\begin{equation}
 W=\cO_V + m Q \tilde{Q}.
\end{equation}
and the theory could have an R-symmetry
depending on the structure of couplings $\cO_V$. Supersymmetry
requires $F_Q = m\, \tilde Q = 0$ and $F_{\tilde Q} = m\, Q = 0$, but
since $U(1)_H$ is light and
the messengers have $U(1)_H$ charge we must also impose the D-term constraint
\begin{equation}
 D=-\xi+ g^2\left( |Q|^2 - |\tilde{Q}|^2 \right)=0,
\end{equation}
where $\xi$ is a non-zero Fayet-Iliopoulos term and $g$ is the gauge
coupling. The scalar potential then takes the form
\begin{equation}
 V = m^2(Q^2+\tilde Q ^2) + \frac{1}{2}\left[ -\xi + g^2(Q^2-\tilde Q ^2) \right]^2,
\end{equation}
and supersymmetry is broken for generic $\xi$ since the D-term and
F-term constraints cannot be simultaneously satisfied, giving $V>0$.
If there is an R-symmetry, taking $\eps$ finite but small generically
introduces a small explicit breaking\footnote{For example, if $\tilde
  \cO_V$ is a particular coupling present in perturbation theory and
  the superpotential has a term $\eps\, \tilde \cO_V Q \tilde Q$, finite but small $\eps$
  gives a small explicit R-symmetry breaking.}, restoring
supersymmetry with a supersymmetric minimum at a distance $\sim
1/\eps$ in field space. The non-supersymmetric vacuum is metastable
for finite $\eps$, perhaps slightly shifted.

\subsubsection{Breaking SUSY with SQCD}

\begin{figure}[t]
 \centering
 \includegraphics[scale=0.7]{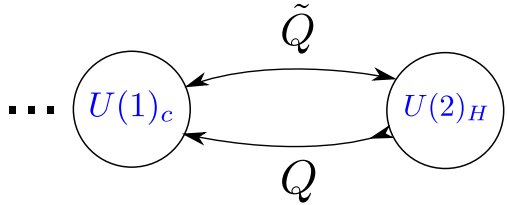}
\caption{A simple realization of supersymmetry breaking via SQCD with messenger flavors.} 
\label{fig:non-madrid-non-abelian}
\end{figure}

In this section we present a quiver where supersymmetric QCD with
messenger flavors breaks supersymmetry as discussed in generality in
section \ref{subsec:susy breaking}. For a concrete example let us take
a $U(N_c)$ node with $N_c=2$ that we call $H$. We take a single
flavor, so we are in the $N_f < N_c$ regime of SQCD. See figure
\ref{fig:non-madrid-non-abelian}.  Messengers are attached to the $c$ node
and the representations are $Q\sim (\ov c,N_c)$ and $\tilde Q \sim
(\ov c, \ov N_c)$.  These are the quarks of SQCD.  These additions
will contribute $T_c = -4$ and therefore this quiver can be added to
any non-Madrid quiver with
$T_c=4\,\text{mod}\,3=1\,\text{mod}\,3$. $U(3N)$ will never cancel $T_c$ charge.

Though $Q\tilde Q$ carries anomalous $U(1)$ charge, it is nevertheless
possible \cite{Haack:2006cy} to generate the ADS superpotential. In
addition, the fact that the quarks carry $U(1)_c$ charge does not
influence the SQCD gauge dynamics, since $U(1)_c$ is anomalous and
affects the low energy effective action only through global selection
rules which protect the flavor mass. We assume that flavors obtain
a small mass via instantons or couplings to singlets and that $\cO_V
\cO_M$ couplings can be made small by a parameter $\eps$, perhaps
given by a worldsheet instanton prefactor. At the very least, they are
highly suppressed as in table \ref{table:operators-suppressions}. The
superpotential is given by
\begin{equation}
 W = \cO_V + \eps\, \cO_V \cO_M  + \lambda_1 \frac{\Lambda^{5}_2}{\tilde Q Q } + \lambda_2 \tilde Q Q.
\end{equation}
In the $\eps \rightarrow 0$ and $\lambda_2 \rightarrow 0$ limit, the ADS superpotential breaks supersymmetry but
the potential exhibits runaway behavior. Since $Q \tilde Q$ carries anomalous $U(1)$ charge, it is plausible that there
is an instanton which can generate the messenger mass but is unrelated to the non-perturbative effect which generates the
ADS superpotential. This effect would make $\lambda_2$ finite and lift the runaway.

\section{Conclusions}
\label{sec:conclusions}

In this paper we have studied stringy hidden valley
models. Broadly, these are hidden valleys where the quasi-hidden
sectors are added to the MSSM for the sake of string consistency
conditions which otherwise would not be satisfied. Since it is
notoriously difficult to differentiate between quantum field theories
and string compactifications at low energies, leveraging string
consistency conditions to motivate extensions of the standard model
is an interesting and promising approach. 

We have focused on a subclass of these models heavily motivated by weakly
coupled type II orientifold compactifications and their duals.
We have shown that there are aspects of low energy physics which are
generic in these stringy hidden valleys but are non-generic for an arbitrary
hidden valley. In our models, this is primarily due to the fact that hidden sectors
added for the sake of string consistency have messenger fields which
are chiral under a symmetry $U(1)_V$ in the visible sector. The
consequences are immediate and generic:
\begin{itemize}
\item Messenger masses are protected by symmetry
but can be generated at a low scale by non-perturbative effects
(e.g. D-instantons) or couplings to singlets.
\item Other couplings are highly affected by $U(1)_V$ symmetry. For
  example, couplings of messengers to hidden sector fields are
  forbidden in perturbation theory. See table \ref{table:operators-suppressions}.
\item There are $U(1)_VU(1)_iU(1)_i$
  anomalies which are canceled via the Green-Schwarz mechanism through
  the introduction of Chern-Simons terms  of the form $\phi_V F_i \wedge F_i$
  and $B_V \wedge F_V$. The latter gives a St\" uckelberg mass to the $U(1)_V$ gauge boson.
\item If there is a non-abelian node on which messengers end, there is
  a $U(1)_V SU(N_i) SU(N_i)$ anomaly whose cancellation requires the
  introduction of a $\phi_V Tr(G_i \wedge G_i)$ term.
\item The annihilation cross section of messengers into standard model fields is
  suppressed via the typically large mass of the anomalous $Z'_V$ gauge boson.
\end{itemize}
Heuristically, chirality of the messengers under $U(1)_V$ provides a coupling barrier
between the visible and hidden sector. There is also a ``$Z'_V$ barrier'' which prevents
messenger annihilation into standard model fermions in proportion to the mass of $Z'_V$.

The very broad subclass of stringy hidden valleys where the hidden
sector is ``hypercharged'' exhibits other interesting
features. Splitting the hypercharge embedding as
\begin{equation}
U(1)_Y = U(1)_{Y,V} + U(1)_{Y,H},  
\end{equation}
this subclass is defined as having a non-trivial contribution
$U(1)_{Y,H}$. We call a quiver in this class a ``hidden hypercharge quiver." Generic features
include:
\begin{itemize}
\item Messengers carry hypercharge and contribute to the $U(1)_VYY$
  anomaly coefficient.  Unless precisely canceled by the contribution
  of the visible sector, the chiral spectrum exhibits a $U(1)_VYY$ anomaly. This is always
  the case for extensions of three-node MSSM quivers.
  \item Cancellation of the $U(1)_VYY$ anomaly requires the presence
    of a Chern-Simons term $\phi_V F_Y \wedge F_Y$. This gives rise to
    a $Z'_VB B$ vertex which gives vertices involving $Z'$, $\gamma$,
    and $Z$ after electroweak symmetry breaking.
  \item We require that any hidden hypercharge quiver must
    have hidden sector fields which are MSSM singlets. This requires
    that any non-zero contribution of a hidden sector node to the
    hypercharge must ``propagate'' to any connected hidden sector
    node. Thus, each node in a connected cluster contributes equally
    to the hypercharge embedding, so that the contribution of
    the $m^\text{th}$ cluster uniquely determined by a number
    $q^{(m)}$.
  \item Ensuring that fields in a cluster with $q^{(m)} \ne 0$ are
    MSSM singlets constrains the possible matter content to be two of
    the four possible bifundamental representations, while the other
    two are forbidden, along with symmetric and antisymmetric tensor
    representations.
  \item If $U(1)_{Y,V}$ is massless, then $U(1)_{Y,H}$ is also
    massless by the requirement that $U(1)_Y$ is massless.  This is
    the case of any extension of a three-node MSSM quiver, and gives a
    light $Z'_{Y,H}$ boson in the spectrum.
\end{itemize}
We see that this subclass is greatly simplified and that there are
generically axionic couplings to the hypercharge field strength and
usually an additional light $Z'$ boson. These observations can have
important phenomenological consequences.

We studied possibilities for dark matter and supersymmetry breaking in
light of the rich structure of superpotential couplings, axionic
couplings required for anomaly cancellation, and $Z'$ physics.  
We considered possibilities for messenger dark matter and hidden
sector dark matter. All possibilities are non-baryonic, since string
consistency does not require the addition of messengers charged under
$SU(3)_{QCD}$. Messenger dark matter masses are protected by symmetry
and could remain light. In the common case where $U(1)_{Y,H}$ is
non-trivial and massless, this symmetry could protect messengers from
decay, as could the many other $U(1)$ symmetries.   
Its annihilation cross section into standard model fermions via
$Z'_V$ is suppressed. Messenger dark matter
in extensions of the Madrid embedding has weak interactions. Hidden
sector dark matter candidates are standard model singlets and their
masses are not necessarily protected by symmetry. The possibilities
for messenger dark matter are more constrained than those of hidden
sector dark matter. 
In hidden hypercharge quivers the presence of $U(1)_VYY$ anomalies
requires $Z'_VBB$ vertices and axionic couplings which can give dark matter
annihilation processes with photons in the final state, as suggested in \cite{Dudas:2009uq,Dudas:2012pb,Fan:2012gr} to account
for the tentative  $\gamma$-ray at the Fermi LAT \cite{Weniger:2012tx}.
We emphasize that the annihilation cross section of processes involving an intermediate
anomalous $Z'$ is suppressed by the large $Z'$ mass, would is typically near the GUT
or Planck scale in a string compactification.

We also studied possibilities for the breaking of global supersymmetry
via supersymmetric QCD or a retrofitted Fayet model. There are two
natural possibilities for SQCD breaking: one where the SQCD flavors
are messengers, and another where the flavors are hidden sector
fields. In the case of messenger flavors, their flavor masses are
protected by $U(1)_V$ symmetry and therefore it is plausible
that they are light relative to the SQCD confinement scale. The ADS
superpotential can be generated, despite the fact that
$1/Q\tilde Q$ carries anomalous $U(1)_V$ charge.  Possible values of
$N_f$ and $N_c$ are constrained by string consistency conditions. In
the case of supersymmetry breaking with hidden sector flavors, the
flavors can be vector-like, realizing standard SQCD. However, flavor
masses are typically large and can be integrated out, giving
supersymmetry preserving pure SQCD at low energies. It is also
possible that the hidden sector flavors are chiral, in which case the
SQCD theory is similar to the case of messenger flavors. Supersymmetry
breaking via a Fayet model frequently appears in the limit where some
parameters $\eps$ are small, due to the D-term constraint of the
frequently massless symmetry $U(1)_F$.  In the limit of finite $\eps$
the supersymmetry breaking vacuum becomes metastable. We find it
interesting that there is a natural massless $U(1)$ for realizing the
Fayet model. It is also likely that some stringy hidden valleys
realize the retrofitted O'Raifeartaigh and Polonyi models of
\cite{Aharony:2007db}. Natural possibilities for mediation include
gauge mediation and D-instanton mediation. $Z'$ mediation is also
possible in a subclass of models, though it is not expected to be
dominant.  Minimal gauge mediation cannot be realized in a way which
utilizes the messengers required for string consistency.

We also presented many example quivers realizing the possibilities
discussed in the general section. We discussed three extensions of the
Madrid embedding, which exhibited a $U(1)_iYY$ anomaly,
multiple light $Z'$ bosons, and possible dark matter annihilation
processes with photons in the final state. In two extensions of the non-Madrid
embedding we gave examples of supersymmetry breaking via SQCD and
a Fayet model.

We leave many interesting possibilities for
future work.  One is to uncover other generic features in the class of
models we have proposed. Another is to construct explicit models,
performing a detailed study of new low energy signatures and studying
experimental constraints on the allowed parameter space. It would also
be interesting to study classes of gauge theories arising in other
regions of the string landscape where string consistency conditions
lead to concrete extensions of the standard model; that is, to find other
classes of stringy hidden valleys.

\acknowledgments We thank Shamit Kachru, Gordon Kane, Denis Klevers,
Joseph Marsano, Mark Trodden, and Angel Uranga for useful
conversations related to this work. J.H. thanks Matt Strassler for
conversations on hidden valleys at TASI 2010 and Keith Dienes for
conversations about dark matter. We are particularly grateful to Paul
Langacker for useful conversations, comments on the manuscript, and
recent collaborations. This work is supported in part by DOE grant
DE-SC0007901.  J.H. also acknowledges support from NSF grant
PHY11-25915 and a DOE graduate fellowship. M.C. is also supported by
the Fay R. and Eugene L. Langberg Endowed Chair and the Slovenian
Research Agency (ARRS).

\bibliographystyle{JHEP}
\bibliography{refs}

\end{document}